\documentclass[prd,aps,preprint,showpacs,nofootinbib,superscriptaddress]{revtex4-1}
\usepackage{epsfig}
\usepackage{amsmath}
\usepackage{amssymb}
\usepackage{dsfont}

\usepackage{color}

 \def\Tr{\mbox{Tr}}
 
 \def\Im{\mbox{Im}}

\newcommand{\be}[1]{
\begin{eqnarray}\label{#1}}
\newcommand{\ee}{\end{eqnarray}}

\newcommand{\ud}{\mathrm{d}}
\newcommand{\uL}{\mathcal{L}}

\newcommand{\uTr}{\mathrm{Tr}}

\newcommand{\uslash}{\slash\!\!\!}

\newcommand{\uvec}[1]{\boldsymbol{#1}}

\allowdisplaybreaks

\begin{document}

\preprint{}

\title{Twist-2 Generalized TMDs and the Spin/Orbital Structure \\ of the Nucleon}
\normalsize
\author{K. Kanazawa$^{1,2}$, C. Lorc\'e$^{3}$,
A.~Metz$^{2}$, B. Pasquini$^{4}$, M.~Schlegel$^{5}$
\\
{\normalsize\it  $^1$ Graduate School of Science and Technology, Niigata University, Ikarashi,
Niigata 950-2181, Japan\\
$^2$Department of Physics, Barton Hall,
Temple University, Philadelphia, PA 19122, USA\\
$^3$ IPNO, Universit\'e Paris-Sud, CNRS/IN2P3, 91406 Orsay, France and\\
IFPA, AGO Department, Universit\'e de Li\`ege, Sart-Tilman,
4000 Li\`ege, Belgium \\
$^4$Dipartimento di Fisica, Universit\`{a} degli Studi di Pavia, and \\ 
Istituto Nazionale di Fisica Nucleare, Sezione di Pavia, I-27100 Pavia, Italy\\
$^5$Institute for Theoretical Physics, T\"ubingen University, \\
 Auf der Morgenstelle 14, D-72076 T\"ubingen, Germany}}

\begin{abstract}
Generalized transverse-momentum dependent parton distributions (GTMDs) encode the most general parton structure of hadrons.
Here we focus on two twist-2 GTMDs which are denoted by $F_{1,4}$ and $G_{1,1}$ in parts of the literature.
As already shown previously, both GTMDs have a close relation to orbital angular momentum of partons inside a hadron.
However, recently even the mere existence of $F_{1,4}$  and $G_{1,1}$ has been doubted.
We explain why this claim does not hold.
We support our model-independent considerations by calculating the two GTMDs in the scalar diquark model and in the quark-target model, where we also explicitly check the relation to orbital angular momentum.
In addition, we compute $F_{1,4}$ and $G_{1,1}$ at large transverse momentum in perturbative Quantum Chromodynamics and show that they are nonzero.
\end{abstract}

\maketitle

%
%
\section{Introduction}
\label{s:Intro}
During the past two decades, a lot of attention has been paid to generalized parton distributions (GPDs)
~\cite{Mueller:1998fv,Ji:1996ek,Radyushkin:1996nd,Goeke:2001tz,Diehl:2003ny,Belitsky:2005qn,Boffi:2007yc} and to transverse-momentum dependent parton distributions 
(TMDs)~\cite{Bacchetta:2006tn,D'Alesio:2007jt,Burkardt:2008jw,Barone:2010zz,Aidala:2012mv}. 
Those objects are of particular interest because they describe the three-dimensional parton structure of hadrons --- the distribution of the parton's longitudinal momentum and transverse position in the case of GPDs, and the distribution of the parton's longitudinal momentum and transverse momentum in the case of TMDs.
Even though GPDs and TMDs already are quite general entities, the maximum possible information about the (two-) parton structure of strongly interacting systems is encoded in GTMDs~\cite{Meissner:2008ay,Meissner:2009ww,Lorce:2013pza}.
GTMDs can reduce to GPDs and to TMDs in certain kinematical limits, and therefore they are often denoted as {\it mother distributions}.
The Fourier transform of GTMDs can be considered as Wigner distributions~\cite{Ji:2003ak,Belitsky:2003nz,Lorce:2011kd}, the quantum-mechanical analogue of classical phase-space distributions.
 
A classification of GTMDs for quarks (through twist-4) for a spin-0 target was given in Ref.~\cite{Meissner:2008ay}, followed by a corresponding work for a spin-$\tfrac{1}{2}$ target~\cite{Meissner:2009ww}.
In Ref.~\cite{Lorce:2013pza}, the counting of quark GTMDs was independently confirmed, and a complete classification of gluon GTMDs was provided as well.
GTMDs were computed in spectator models~\cite{Meissner:2008ay,Meissner:2009ww}, and in two different light-front quark models~\cite{Lorce:2011dv,Lorce:2011kd}.
Gluon GTMDs quite naturally appear when describing high-energy diffractive processes like vector meson production or Higgs production in so-called $k_T$-factorization in Quantum Chromodynamics (QCD)~\cite{Martin:1999wb,Khoze:2000cy,Martin:2001ms}.
It is important to note that, in general, there exists no argument according to which, as a matter of principle, GTMDs cannot be measured, even though for quark GTMDs a proper high-energy process has yet to be identified.  

Recent developments revealed an intimate connection between two specific GTMDs --- denoted by $F_{1,4}$ and $G_{1,1}$ in Ref.~\cite{Meissner:2009ww} --- and the spin/orbital structure of the nucleon.
(See Refs.~\cite{Leader:2013jra,Wakamatsu:2014zza} for recent reviews on the decomposition of the nucleon spin.)
In particular, the relation between $F_{1,4}$ and the orbital angular momentum (OAM) of partons inside a longitudinally polarized nucleon~\cite{Lorce:2011kd,Hatta:2011ku,Lorce:2011ni,Ji:2012sj,Lorce:2012ce} has already attracted considerable attention.
As shown in~\cite{Lorce:2011kd}, this relation gets its most intuitive meaning when expressed in terms of the Wigner function $\mathcal{F}_{1,4}$, i.e., the Fourier transform of $F_{1,4}$.
It is also quite interesting that, depending on how one chooses the path of the gauge link that makes the GTMD correlator gauge invariant~\cite{Ji:2012sj,Lorce:2012ce}, $F_{1,4}$ provides either the (canonical) OAM of Jaffe-Manohar~\cite{Jaffe:1989jz} or the OAM in the definition of Ji~\cite{Ji:1996ek}.
(We also refer to~\cite{Burkardt:2012sd} for a closely related discussion.)
The connection between OAM and $F_{1,4}$ could make the canonical OAM accessible to Lattice QCD as first pointed out in~\cite{Hatta:2011ku}.
The GTMD $G_{1,1}$ can be considered as ``partner" of $F_{1,4}$ as it describes longitudinally polarized partons in an unpolarized nucleon~\cite{Lorce:2011kd,Lorce:2014mxa}.
The very close analogy between those two functions becomes most transparent if one speaks in terms of spin-orbit correlations~\cite{Lorce:2014mxa}.
While $F_{1,4}$ quantifies the correlation between the nucleon spin and the OAM of partons, $G_{1,1}$ quantifies the correlation between the parton spin and the OAM of partons~\cite{Lorce:2014mxa}.

Despite all those developments, recently it has been argued that OAM of partons cannot be explored through leading-twist GTMDs~\cite{Liuti:2013cna,Courtoy:2013oaa}.
In fact, the mere existence of $F_{1,4}$  and $G_{1,1}$ has been doubted.
Here we refute this criticism and explain why the arguments given in~\cite{Liuti:2013cna,Courtoy:2013oaa} do not hold.
A key problem of the work in~\cite{Liuti:2013cna,Courtoy:2013oaa} is the application of a two-body scattering picture for the classification of GTMDs, which turns out to be too restrictive. 
In order to make as clear as possible that in general neither $F_{1,4}$ nor $G_{1,1}$ vanish, we also compute them in the scalar diquark model of the nucleon, in the quark-target model, and in perturbative QCD for large transverse parton momenta.
In addition, for the two models, we show explicitly the connection between the two GTMDs and the OAM of partons.
The model calculations are carried out in standard perturbation theory for correlation functions, and also by using the overlap representation in terms of light-front wave functions (LFWFs) where some of the results become rather intuitive.

The paper is organized as follows:
In Sect.~\ref{s:GTMDs} we specify our conventions and repeat some key ingredients for the counting of GTMDs, while Sect.~\ref{s:Two-body} essentially deals with our reply to the arguments given in~\cite{Liuti:2013cna,Courtoy:2013oaa}.
The calculations of $F_{1,4}$ and $G_{1,1}$ in the scalar diquark model and in the quark-target model are discussed in Sect.~\ref{s:Models}, along with their relation to OAM of partons.
In Sect.~\ref{s:Overlap} we repeat the studies of Sect.~\ref{s:Models} by making use of LFWFs.
The computation of the two GTMDs at large transverse momentum is presented in Sect.~\ref{s:PQCD}.
We summarize the paper in Sect.~\ref{s:Summary}.

%
%
\section{Definition of GTMDs}
\label{s:GTMDs}
{In this section we review the definition of the GTMDs for quarks, which have been classified in Refs.~\cite{Meissner:2008ay,Meissner:2009ww} and further discussed also for the gluon sector in Ref.~\cite{Lorce:2013pza}.
We begin by introducing two light-like four-vectors $n_\pm$ satisfying $n_\pm^2 = 0$ and $n_+\cdot n_-=1$.
They allow one to} decompose a generic four-vector $a^\mu$ as
\begin{equation}
 a=a^+n_++a^-n_-+a_\perp,
\end{equation}
where the transverse light-front four-vector is defined as 
$a^\mu_\perp=g_\perp^{\mu\nu}a_\nu$, with $g_\perp^{\mu\nu}=g^{\mu\nu}-n_+^\mu n_-^\nu-n_-^\mu n_+^\nu$.
We consider the following fully-unintegrated quark-quark correlator  
\begin{equation}\label{qGPCF}
W^{[\Gamma]}_{\Lambda'\Lambda}(P,\bar k,\Delta,n_-;\eta)=\frac{1}{2}\int\frac{\ud^4z}{(2\pi)^4}\,e^{i \bar k\cdot z}\,\langle p',\Lambda'|T\{\overline\psi(-\tfrac{z}{2})\Gamma\,\mathcal W_{n_-}\psi(\tfrac{z}{2})\}|p,\Lambda\rangle,
\end{equation}
which depends on the initial (final) hadron light-front helicity $\Lambda$ ($\Lambda'$), and the following independent four-vectors: 
\begin{itemize}
\item the average nucleon four-momentum $P=\tfrac{1}{2}(p'+p)$;
\item the average quark four-momentum $\bar k=\tfrac{1}{2}(k'+k)$;
\item the four-momentum transfer $\Delta=p'-p=k'-k$;
\item the light-like four-vector $n_-$.
\end{itemize}
The object $\Gamma$ in Eq.~\eqref{qGPCF} stands  for any element of the basis $\{\mathds 1,\gamma_5,\gamma^\mu,\gamma^\mu\gamma_5,i\sigma^{\mu\nu}\gamma_5\}$ in Dirac space. 
The Wilson contour  $\mathcal W_{n_-}\equiv\mathcal W(-\tfrac{z}{2},\tfrac{z}{2}|\eta n_-)$ ensures the color gauge invariance of the correlators,
connecting the points $-\tfrac{z}{2}$ and $\tfrac{z}{2}$ \emph{via} the intermediate points $-\tfrac{z}{2}+\eta\infty n_-$ and $\tfrac{z}{2}+\eta\infty n_-$. 
The parameter $\eta=\pm$ indicates whether the Wilson contour is future-pointing or past-pointing. 
The four-vector $n_-$ defines the light-front direction which allows one to organize the structure of the correlator as an expansion in $\left(\tfrac{M}{P^+}\right)^{t-2}$, where $t$ defines the operational twist~\cite{Jaffe:1996zw}. 
Without loss of generality, we choose the $z$-axis along the  $\vec n_+=-\vec n_-$ direction, i.e.,  $n_-=\tfrac{1}{\sqrt{2}}(1,0,0,-1)$ and $n_+=\tfrac{1}{\sqrt{2}}(1,0,0,1)$.
Therefore, the light-front coordinates of a generic four-vector $a=[a^+,a^-,\uvec a_\perp]$ are given by $a^+=\tfrac{1}{\sqrt{2}}(a^0+a^3)$, $a^-=\tfrac{1}{\sqrt{2}}(a^0-a^3)$, and $\uvec a_\perp=(a^1,a^2)$.

Since the light-front quark energy is hard to measure, one can focus on the $\bar k^-$-integrated version of Eq.~\eqref{qGPCF},
\begin{equation}\label{qGTMD}
W^{[\Gamma]}_{\Lambda'\Lambda}(P,x,\bar{\uvec k}_\perp,\Delta,n_-;\eta)=\frac{1}{2}\int\ud\bar k^-\!\int\frac{\ud^4z}{(2\pi)^4}\,e^{i \bar k\cdot z}\,\langle p',\Lambda'|\overline\psi(-\tfrac{z}{2})\Gamma\,\mathcal W_{n_-}\psi(\tfrac{z}{2})|p,\Lambda\rangle,
\end{equation}
where time-ordering is not needed anymore and $\bar k=[xP^+,\bar k^-,\bar{\uvec k}_\perp]$. 
The functions that parametrize this correlator are called GTMDs and can be seen as the {\it mother distributions} of GPDs and TMDs~\cite{Meissner:2008ay,Meissner:2009ww,Lorce:2013pza}.

\subsection{Helicity Amplitudes and GTMD Counting}
\label{subs:GTMDcounting}
In the following, we will restrict our discussion to leading twist $t=2$, and we refer to \cite{Meissner:2009ww,Lorce:2013pza} for a full analysis up to twist 4. 
In the region $x>\xi$, with $\xi=-\frac{\Delta^+}{2P^+}$, where the correlator \eqref{qGTMD} describes the emission of a quark with momentum $k$ and helicity $\lambda$ from the nucleon and its reabsorption with momentum $k'$ and helicity $\lambda'$, it is convenient to introduce light-front helicity amplitudes~\cite{Diehl:2001pm} according to
\begin{equation}\label{HelicityAmpl}
H_{\Lambda'\lambda',\Lambda\lambda}(P,x,\bar{\uvec k}_\perp,\Delta,n_-;\eta)=\langle p',\Lambda'|  \mathcal O_{\lambda'\lambda}(x,\bar{\uvec k}_\perp,n_-;\eta)|p,\Lambda\rangle,
\end{equation}
where in the chiral-even sector
\begin{equation}
 \mathcal O_{\pm\pm}(x,\bar{\uvec k}_\perp,n_-;\eta)=\frac{1}{2}\int\ud\bar k^-\!\int\frac{\ud^4z}{(2\pi)^4}\,e^{i \bar k\cdot z}\,\overline\psi(-\tfrac{z}{2})\,\gamma^+(\mathds 1\pm\gamma_5)\,\mathcal W_{n_-}\psi(\tfrac{z}{2}),
\end{equation}
and in the chiral-odd sector
\begin{equation}
 \mathcal O_{\pm\mp}(x,\bar{\uvec k}_\perp,n_-;\eta)=\frac{1}{2}\int\ud\bar k^-\!\int\frac{\ud^4z}{(2\pi)^4}\,e^{i \bar k\cdot z}\,\overline\psi(-\tfrac{z}{2})\,i\sigma^{R(L)+}\gamma_5\,\mathcal W_{n_-}\psi(\tfrac{z}{2}),
\end{equation}
with $a^{R(L)}=a^1\pm ia^2$.

The light-front discrete symmetry\footnote{The light-front parity and time-reversal transformations consist of the ordinary parity and time-reversal transformations followed by a $\pi$-rotation about the $y$-axis~\cite{Soper:1972xc,Carlson:2003je,Brodsky:2006ez}. Contrary to the ordinary discrete transformations, the light-front ones preserve the light-front vector $n_-$.} and hermiticity constraints imply relations among these light-front helicity amplitudes~\cite{Lorce:2013pza}:
\begin{align}
\text{Hermiticity}\quad H_{\Lambda'\lambda',\Lambda\lambda}(P,x,\bar{\uvec k}_\perp,\Delta,n_-;\eta)&=H^*_{\Lambda\lambda,\Lambda'\lambda'}(P,x,\bar{\uvec k}_\perp,-\Delta,n_-;\eta),\\
\text{LF Parity}\quad H_{\Lambda'\lambda',\Lambda\lambda}(P,x,\bar{\uvec k}_\perp,\Delta,n_-;\eta)&=H_{-\Lambda'-\lambda',-\Lambda-\lambda}(P_{\mathsf P},x,\bar{\uvec k}_{\perp\mathsf P},\Delta_{\mathsf P},n_-;\eta),
\label{eq:parity-hel}\\
\text{LF Time-reversal}\quad H_{\Lambda'\lambda',\Lambda\lambda}(P,x,\bar{\uvec k}_\perp,\Delta,n_-;\eta)&=(-1)^{\Delta\ell_z}H^*_{\Lambda'\lambda',\Lambda\lambda}(P_{\mathsf P},x,\bar{\uvec k}_{\perp\mathsf P},\Delta_{\mathsf P},n_-;-\eta),\label{TR-hel}
\end{align}
where $a_{\mathsf P}=[a^+,a^-, \uvec a_{\perp\mathsf P}]$ with $\uvec a_{\perp\mathsf P}=(-a^1,a^2)$, and $\Delta\ell_z=(\Lambda-\lambda)-(\Lambda'-\lambda')$. 
\newline

There are 16 helicity configurations, but light-front parity reduces the number of the independent ones to 8. 
As discussed hereafter, each of these amplitudes can be parametrized in terms of two independent Lorentz structures multiplied by scalar functions $X$, leading to a total of 16 independent functions, known as GTMDs. 
This counting agrees with the conclusion drawn in Ref.~\cite{Meissner:2009ww} based on a different but equivalent manifestly covariant approach.

Let us now explain why there are two GTMDs associated with each of the 8 independent helicity amplitudes. 
The GTMDs can only depend on variables that are invariant under the transformations which preserve the light-front vector $n_-$ up to a scaling factor, namely the kinematic Lorentz transformations (the three light-front boosts and the rotations about the $z$-axis) and the light-front parity transformation\footnote{For convenience, we do not include light-front time-reversal at this stage. The counting then leads to 16 \emph{complex-valued} GTMDs with no definite transformation properties under light-front time-reversal. Including light-front time-reversal gives at the end 32 \emph{real-valued} functions with definite transformation properties. The parametrizations in Refs.~\cite{Meissner:2009ww,Lorce:2013pza} have been chosen such that the real part of the GTMDs is (naive) $\mathsf T$-even and the imaginary part is (naive) $\mathsf T$-odd.}. 
From the independent four-vectors at our disposal, these variables are\footnote{In general the GTMDs depend also on $\eta$.} $(x,\xi,\uvec\kappa^2_\perp,\uvec \kappa_\perp\cdot\uvec D_\perp,\uvec D^2_\perp)$, where $\uvec\kappa_\perp=\bar{\uvec k}_\perp-x\uvec P_\perp$ and $\uvec D_\perp=\uvec\Delta_\perp+2\xi\uvec P_\perp$. 
By conservation of the total angular momentum along the $z$-direction, each light-front helicity amplitude is associated with a definite OAM transfer $\Delta\ell_z$. 
Since there are only two frame-independent (or intrinsic) transverse vectors available, the general structure of a light-front helicity amplitude is given in terms of explicit global powers of $\uvec\kappa_\perp$ and $\uvec D_\perp$, accounting for the OAM transfer, multiplied by a scalar function $X(x,\xi,\uvec\kappa^2_\perp,\uvec \kappa_\perp\cdot\uvec D_\perp,\uvec D^2_\perp;\eta)$. 
As explicitly discussed in Ref.~\cite{Lorce:2013pza}, from two independent transverse vectors, one can form only two independent Lorentz structures associated with a given $\Delta\ell_z$, or, equivalently, light-front helicity amplitude.

\subsection{GTMD Parametrization}
\label{subs:GTMDpara}
Since the GTMD counting is frame-independent, we are free to work in any frame. For convenience, we choose the symmetric frame where the momenta are given by
\begin{equation}\label{symmetricframe}
\begin{aligned}
 P&=\left[P^+,P^-,\uvec 0_\perp\right ],\\
 \bar k&=\left[xP^+,\bar k^-,\bar{\uvec k}_\perp \right],\\
\Delta&=\left[-2\xi P^+,2\xi P^-,\uvec\Delta_\perp\right]
\end{aligned}
\end{equation}
with $P^-=\frac{M^2+\uvec \Delta_\perp^2/4}{2(1-\xi^2)P^+}$,  and the intrinsic transverse vectors simply reduce to $\uvec\kappa_\perp=\bar{\uvec k}_\perp$ and $\uvec D_\perp=\uvec\Delta_\perp$.
Restricting ourselves to the chiral-even sector and using the parametrization in Ref.~\cite{Meissner:2009ww}, the light-front helicity amplitudes for $\xi=0$,  $\Lambda'=\Lambda=\pm$ and  $\lambda'=\lambda=\pm$ read
\begin{equation}\label{helicity}
H_{\Lambda\lambda,\Lambda\lambda}=\tfrac{1}{2}\left[F_{1,1}+\Lambda\lambda\,G_{1,4}+\tfrac{i(\bar{\uvec k}_\perp\times\uvec\Delta_\perp)_z}{M^2}\left(\Lambda\,F_{1,4}-\lambda\,G_{1,1}\right)\right].
\end{equation}
Inverting this expression, we obtain for $F_{1,4}$ and $G_{1,1}$
\begin{align}
\tfrac{i(\bar{\uvec k}_\perp\times\uvec\Delta_\perp)_z}{M^2}\,F_{1,4}&=\tfrac{1}{2}\left[H_{++,++}+H_{+-,+-}-H_{-+,-+}-H_{--,--}\right],\label{F14ampl}\\
-\tfrac{i(\bar{\uvec k}_\perp\times\uvec\Delta_\perp)_z}{M^2}\,G_{1,1}&=\tfrac{1}{2}\left[H_{++,++}-H_{+-,+-}+H_{-+,-+}-H_{--,--}\right].\label{G11ampl}
\end{align}
The function $F_{1,4}$ describes how the longitudinal polarization of the target distorts the unpolarized distribution of quarks, whereas $G_{1,1}$ describes how the longitudinal polarization of quarks distorts their distribution inside an unpolarized target~\cite{Lorce:2011kd}.

%
%
\section{The two-body scattering picture}
\label{s:Two-body}
We are now in a position to address the criticism put forward in Refs.~\cite{Liuti:2013cna,Courtoy:2013oaa}. 
In these papers it has been argued that, based on parity arguments in a two-body scattering picture, the functions $F_{1,4}$ and $G_{1,1}$ should not be included in a twist-2 parametrization.
This contradicts the findings of Refs.~\cite{Meissner:2008ay,Meissner:2009ww,Hatta:2011ku,Lorce:2013pza} and the results obtained in explicit model calculations~\cite{Meissner:2008ay,Meissner:2009ww,Lorce:2011kd,Lorce:2011ni}.
In the following, we will go along the arguments developed in Refs.~\cite{Liuti:2013cna,Courtoy:2013oaa} and explain why they actually do not hold. 
We note essentially three claims in Refs.~\cite{Liuti:2013cna,Courtoy:2013oaa} :
\begin{enumerate}
 \item  The Lorentz structure associated with the function $F_{1,4}$ and appearing in the parametri\-zation of the correlator $W^{[\gamma^+]}_{\Lambda'\Lambda}$ in Ref.~\cite{Meissner:2009ww}
\begin{equation}\label{F14struc}
\overline u(p',\Lambda')\,\frac{i\sigma^{jk}\bar k^j_\perp\Delta^k_\perp}{M^2}\,u(p,\Lambda)\propto\langle\vec S_L\cdot(\vec{\bar k}_\perp\times\vec\Delta_\perp)\rangle
\end{equation}
is parity-odd. 
In the center-of-mass (CM) frame, or equivalently in the ``lab'' frame, with the $p$-direction chosen as the $z$-direction, the net longitudinal polarization defined in Eq.~\eqref{F14struc} is clearly a parity violating term (pseudoscalar) under space inversion. 
This implies that a measurement of single longitudinal polarization asymmetries would violate parity conservation in an ordinary two-body scattering process corresponding to tree-level, twist-2 amplitudes.
\item As one can see from Eqs.~\eqref{F14ampl} and \eqref{G11ampl}, the functions $F_{1,4}$  and $G_{1,1}$ can be nonzero only when the corresponding helicity amplitude combinations are imaginary. 
Hence, these functions cannot have a straightforward partonic interpretation. Moreover, integrating e.g.~Eq.~\eqref{F14ampl} over $\bar{\uvec k}_\perp$ gives zero, meaning that this term decouples from partonic angular momentum sum rules.
\item In the CM frame, where the hadron and quark momenta are coplanar, there must be another independent direction for the helicity amplitude combinations associated with $F_{1,4}$ and $G_{1,1}$ to be non-zero. 
That is provided by twist-3 amplitudes and corresponding GTMDs.
\end{enumerate}

Let us first discuss the parity property of the Lorentz structure in Eq.~\eqref{F14struc}. 
Parity is a frame-independent symmetry, and so does not depend on any particular frame. 
Under an ordinary parity transformation, see for instance Chapter 3.6 of Ref.~\cite{Peskin:1995ev}, the structure in Eq.~\eqref{F14struc} becomes
\begin{equation}\label{F14Pstruc}
\overline u(p'_{\mathtt P},\Lambda'_{\mathtt P})\,\frac{i\sigma^{jk}\bar k^j_\perp\Delta^k_\perp}{M^2}\,u(p_{\mathtt P},\Lambda_{\mathtt P}),
\end{equation}
where $\Lambda_{\mathtt P}$ and $p_{\mathtt P}=(p^0,-\vec p\,)$ are the parity-transformed helicity and momentum. 
Clearly, the structure \eqref{F14struc} does not change sign under a parity transformation, and so is parity-even\footnote{Note also that the structure \eqref{F14struc} has no uncontracted index, implying that its parity coincides with its intrinsic parity. Since none of the only objects with odd intrinsic parity (\emph{i.e.} $\epsilon_{\mu\nu\rho\sigma}$ and $\gamma_5$) is involved, the structure \eqref{F14struc} is automatically parity-even.}. 
This is consistent with the light-front helicity amplitudes given in Eq.~\eqref{helicity}, where the structure \eqref{F14struc} contributes in the form $\Lambda\,\tfrac{i(\bar{\uvec k}_\perp\times\uvec\Delta_\perp)_z}{M^2}$ which is invariant under light-front parity transformation. 
While it is true that the net longitudinal polarization $S_L=\vec S\cdot\vec P/|\vec P|$ is parity-odd and therefore cannot generate 
by itself a parity-conserving single-spin asymmetry, it is actually multiplied in Eq.~\eqref{F14struc} by $(\bar{\uvec k}_\perp\times\uvec\Delta_\perp)_z=(\vec{\bar k}\times\vec \Delta)\cdot\vec P/|\vec P|$, another parity-odd quantity which is related to the quark OAM. 
Therefore, the parity argument cannot be used to exclude the possible connection between a single-spin asymmetry and the function $F_{1,4}$. 
The same is true for $G_{1,1}$ using a similar argument where the nucleon polarization is basically replaced by the quark polarization.
We agree that longitudinal single-spin effects for elastic two-body scattering are necessarily parity-violating, but the two-body picture in general cannot be used for the counting of GTMDs as we explain in more detail below.

Now we move to the second claim.
Contrary to TMDs and GPDs, the GTMDs are complex-valued functions. 
The two-body scattering picture does not incorporate any initial or final-state interactions which, in particular, generate (naive) $\mathsf T$-odd effects described by the imaginary parts of the GTMDs.
This already shows that two-body scattering could at best be applied to the real part of the GTMDs which is (naive) $\mathsf T$-even.
The fact that the real parts of $F_{1,4}$ and $G_{1,1}$ are related to imaginary helicity amplitude combinations does not necessarily mean that these GTMDs cannot have a straightforward partonic interpretation.  
The same occurs in the GPD case, for instance, where complex-valued helicity amplitudes do not spoil a partonic interpretation of GPDs (see e.g.~Ref.~\cite{Diehl:2003ny}).
This can also be understood from the partonic interpretation of the distributions in impact-parameter space.
As shown in Ref.~\cite{Lorce:2011kd},  which generalizes the work of Burkardt~\cite{Burkardt:2000za,Burkardt:2002hr} to phase-space, the  GTMDs in impact-parameter space are obtained from the momentum-transfer space through the following two-dimensional Fourier transform
\begin{equation}
\mathcal X(x,\bar{\uvec k}_\perp,\uvec b_\perp)=\int\frac{\ud^2{\uvec \Delta}_\perp}{(2\pi)^2}\,e^{-i\uvec\Delta_\perp\cdot\uvec b_\perp}\, X(x,\xi=0,\bar{\uvec k}_\perp,\uvec\Delta_\perp).
\end{equation}
The hermiticity property of the GTMD correlator guarantees that the two-dimensional Fourier transform is real~\cite{Lorce:2011ni}. 
In particular, the impact-parameter space versions of $F_{1,4}$ and $G_{1,1}$ follow from the two-dimensional Fourier transform of Eqs.~\eqref{F14ampl} and \eqref{G11ampl},
\begin{align}
-\tfrac{(\bar{\uvec k}_\perp\times\partial_{\uvec b_\perp})_z}{M^2}\,\mathcal F_{1,4}&=\tfrac{1}{2}\left[\mathcal H_{++,++}+\mathcal H_{+-,+-}-\mathcal H_{-+,-+}-\mathcal H_{--,--}\right],\\
\tfrac{(\bar{\uvec k}_\perp\times\partial_{\uvec b_\perp})_z}{M^2}\,\mathcal G_{1,1}&=\tfrac{1}{2}\left[\mathcal H_{++,++}-\mathcal H_{+-,+-}+\mathcal H_{-+,-+}-\mathcal H_{--,--}\right].
\end{align}
In impact-parameter space, the helicity amplitude combinations associated with $\mathcal F_{1,4}$ and $\mathcal G_{1,1}$ are real, and so are suitable for a partonic interpretation. 
Also, since the GTMDs are functions of the transverse momenta \emph{via} the scalar combinations $\bar{\uvec k}^2_\perp$, $\bar{\uvec k}_\perp\cdot\uvec\Delta_\perp$ and $\uvec\Delta^2_\perp$,  Eqs.~\eqref{F14ampl} and \eqref{G11ampl} give automatically zero in the limit $|\uvec\Delta_\perp|=0$ or by integration over $\bar{\uvec k}_\perp$, implying that $F_{1,4}$ and $G_{1,1}$ do not reduce to any GPD or TMD. 
However, this does not mean that $F_{1,4}$ decouples from partonic angular momentum sum rules, since it has been shown independently in Refs.~\cite{Lorce:2011kd} and \cite{Hatta:2011ku} that the (gauge-invariant) quark canonical OAM of Jaffe-Manohar~\cite{Jaffe:1989jz} is actually given by the expression
\begin{align}
\ell^q_z&=\int\ud x\,\ud^2\bar {\uvec k}_\perp \, \ud^2 {\uvec b}_\perp\,(\uvec b_\perp\times\bar{\uvec k}_\perp)_z\,\left[-\tfrac{(\bar{\uvec k}_\perp\times\partial_{\uvec b_\perp})_z}{M^2}\,\mathcal F_{1,4}\right]\nonumber\\
&=\int\ud x\,\ud^2\bar {\uvec k}_\perp\,\,(\bar{\uvec k}_\perp\times i\partial_{\uvec\Delta_\perp})_z\left[\tfrac{i(\bar{\uvec k}_\perp\times\uvec\Delta_\perp)_z}{M^2}\,F_{1,4}\right]_{|\uvec\Delta_\perp|=0}\nonumber\\
&=-\int\ud x\,\ud^2\bar {\uvec k}_\perp\,\tfrac{\bar{\uvec k}_\perp^2}{M^2}\,F_{1,4}(x,0,\bar{\uvec k}^2_\perp,0,0;\eta),\label{OAMformula}
\end{align}
which \emph{a priori} has no reason to vanish. 
Similarly, the amplitude associated with the GPD $E$ vanishes in the limit $|\uvec\Delta_\perp|=0$, but this does not mean that $E$ should disappear from the Ji relation for angular momentum~\cite{Ji:1996ek}.

\begin{figure}[t!]
\begin{center}
\epsfig{file=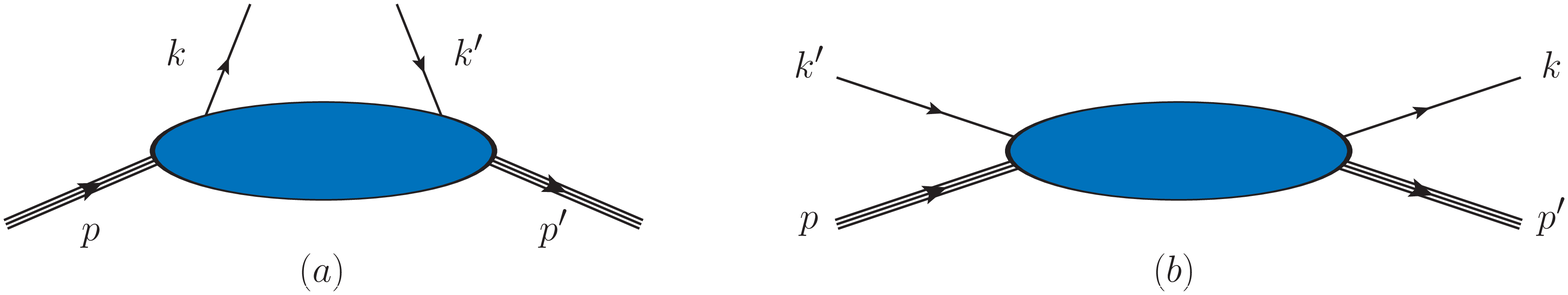,  width=\columnwidth}
\end{center}
\caption{\footnotesize{(a) Representation of a GTMD in the region $\xi<x<1$;
(b) Quark-proton scattering amplitude.}}
\vspace{-0.2 truecm}
\label{fig1}
\end{figure}
\begin{figure}[t!]
\begin{center}
\epsfig{file=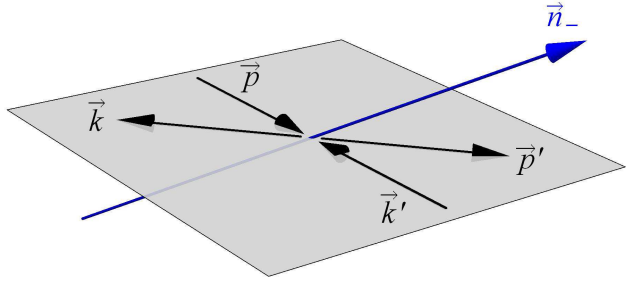,  width=.5\textwidth}
\end{center}
\caption{\footnotesize{Two-body elastic scattering picture in the center-of-mass frame. 
In general, the vector $\vec n_-$ does not belong to the scattering plane.}}
\vspace{-0.2 truecm}
\label{fig2}
\end{figure}

Finally, we address the third claim. 
The helicity amplitudes have been used several times in the literature for the counting of independent functions associated with a particular parton-parton correlator, see for instance Refs.~\cite{Jaffe:1996zw,Hoodbhoy:1998vm,Diehl:2001pm,Lorce:2013pza}. 
It is often considered convenient to think of the correlator in Eq.~\eqref{qGTMD} as representing a two-body elastic scattering amplitude, see Fig.~\ref{fig1}. 
The gauge link is excluded from this picture, and so are initial and final-state interactions, which eliminates also the $\eta$ dependence.
The light-front parity constraint (\ref{eq:parity-hel}) can then be rewritten in the form
\begin{equation}
H_{-\Lambda'-\lambda',-\Lambda-\lambda}(P,x,\bar{\uvec k}_\perp,\Delta,n_-)=(-1)^{\Delta\ell_z}H^*_{\Lambda'\lambda',\Lambda\lambda}(P,x,\bar{\uvec k}_\perp,\Delta,n_-).
\end{equation}
So, from the 16 possible helicity amplitudes only 8 are independent, in agreement with our general discussion in Sect.~\ref{subs:GTMDcounting}. 
Although the counting is frame-independent, the authors of Refs.~\cite{Liuti:2013cna,Courtoy:2013oaa} chose to work for convenience in the CM frame, or equivalently the lab frame, where all momenta are coplanar.
If one chooses the $z$-axis to lie in the scattering plane like in Refs.~\cite{Liuti:2013cna,Courtoy:2013oaa}, one would then conclude that there is only one frame-independent or intrinsic transverse vector, and so the cross product $(\bar{\uvec k}_\perp\times\uvec\Delta_\perp)_z$ would simply give zero. 
For chiral-even, non-flip amplitudes, instead of the four functions $F_{1,1}$, $F_{1,4}$, $G_{1,1}$ and $G_{1,4}$ introduced in Ref.~\cite{Meissner:2009ww}, only $F_{1,1}$ and $G_{1,4}$ would survive. 
However, as stressed by Diehl in Ref.~\cite{Diehl:2001pm}, the two-body scattering formulation is somewhat imprecise. The $H_{\Lambda'\lambda',\Lambda\lambda}$ are not helicity amplitudes in the strict sense. 
They contain, in particular, an explicit dependence on the light-front vector $\vec n_-$ which already defines the $z$-direction\footnote{For a given scattering process, the four-vector $n_-$ 
can be defined in terms of the physically relevant four-vectors. This vector is independent of the three four-vectors one has in the quark-nucleon scattering picture. In the case of deep-inelastic scattering, for instance, it is the four-momentum of the exchanged virtual gauge boson which provides another independent four-vector. After factorization, even ordinary forward parton distributions still know about $n_-$.}. 
One cannot perform an arbitrary Lorentz transformation that modifies this direction without changing at the same time the definition of the GTMDs, and hence the canonical twist expansion.
In the most general configuration, $\vec n_-$ does not belong to the CM scattering plane (see Fig.~\ref{fig2}). 
Therefore, there are \emph{two} independent intrinsic transverse vectors at leading twist, namely $\uvec\kappa_\perp$ and $\uvec D_\perp$, allowing one to form the cross-product $(\uvec\kappa_\perp\times\uvec D_\perp)_z$ which reduces to $(\bar{\uvec k}_\perp\times\uvec\Delta_\perp)_z$ in the symmetric frame \eqref{symmetricframe} used for the parametrization of Ref.~\cite{Meissner:2009ww}. 
One does not need to invoke higher-twist to provide the missing direction. 
It is essential to keep the explicit $n_-$ dependence, since otherwise one would miss half of the allowed GTMDs, just like one would obtain two chiral-odd GPDs~\cite{Hoodbhoy:1998vm} instead of four~\cite{Diehl:2001pm}. 
Note also that applying the two-body scattering picture without $n_-$ dependence to the TMD correlator, one would find only three independent (naive) ${\mathsf T}$-even functions $f_1(x,\bar{\uvec k}^2_\perp)$, $g_{1L}(x,\bar{\uvec k}^2_\perp)$ and $h_1(x,\bar{\uvec k}^2_\perp)$ instead of six, since $\bar{\uvec k}_\perp=-\uvec\Delta_\perp=\uvec 0_\perp$ in the CM frame with $\uvec P_\perp=\uvec 0_\perp$.

%
%
\section{Perturbative model results}
\label{s:Models}
After the model-independent considerations in the previous two sections we now proceed to discuss the calculation of $F_{1,4}$ and $G_{1,1}$ in two spectator models --- the scalar diquark model and the quark-target model.
In fact, in the (parity-conserving) scalar diquark model nonzero results for those two GTMDs were already presented in Ref.~\cite{Meissner:2009ww}~using a diagrammatic approach, but in view of the doubts raised in~\cite{Liuti:2013cna,Courtoy:2013oaa} we reconsider the calculation here also in the context of light-front quantization.
Moreover, a set of GTMDs including $F_{1,4}$ and $G_{1,1}$ was calculated in two different relativistic light-front quark models in Ref.~\cite{Lorce:2011dv,Lorce:2011kd}, where it also turned out that both $F_{1,4}$ and $G_{1,1}$ are non-vanishing.
In Ref.~\cite{Liuti:2013cna} it is argued that ``these nonzero results are coming about from the kinematics or from effective higher-twist components arising from quarks' confinement''. 
We disagree with this statement. 
First, kinematics cannot be invoked to explain nonzero results of
existing calculations~\cite{Lorce:2011kd,Lorce:2011dv}, since they were performed using the
representation of the GTMDs in terms of LFWFs which are by
definition/construction frame-independent.
Moreover, using light-front quantization, the quark correlation functions~\eqref{HelicityAmpl} entering the calculation of $F_{1,4}$ and $G_{1,1}$ have a decomposition which involve only the ``good'' light-front components of the fields, and therefore correspond to pure  twist-two contributions (see, for example, Ref.~\cite{Jaffe:1996zw}).
In this context note also that in spectator models the active partons are (space-like) off-shell. 
However, this feature, which is not specific to the calculation of GTMDs but rather holds even for ordinary forward parton distributions, does not increase the counting of independent twist-2 functions.
Second, as we discuss in this section, not only the scalar diquark model but also the quark-target model predicts nonzero results for $F_{1,4}$ and $G_{1,1}$.
Both models being purely perturbative, this means that ``effective higher twist components arising from quarks' confinement'' cannot be invoked to explain these results.

That $F_{1,4}$ and $G_{1,1}$ are nonzero is a generic feature. 
They are directly related to the amount of OAM and spin-orbit correlations inside the target \cite{Lorce:2011kd,Hatta:2011ku,Lorce:2011ni,Ji:2012sj,Lorce:2012ce,Lorce:2014mxa}. 
Vanishing $F_{1,4}$ and $G_{1,1}$ would therefore imply vanishing (canonical) OAM and spin-orbit correlations. 
In order to further solidify the relation between $F_{1,4}$ and $G_{1,1}$ and the spin/orbital structure of the nucleon, we compute the canonical OAM and spin-orbit correlations from the operator definition in both the scalar diquark model and the quark-target model. 
They are nonzero and satisfy the model-independent relations of Refs.~\cite{Lorce:2011kd,Hatta:2011ku,Lorce:2011ni,Lorce:2014mxa}

For convenience, we will restrict the discussions in the following to the case $\xi=0$.

\subsection{Scalar Diquark Model}
\label{subs:SDM}
\begin{figure}[t!]
\begin{center}
\epsfig{file=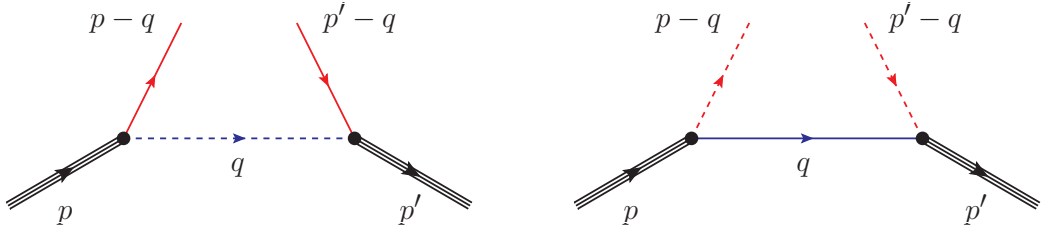,  width=.85\textwidth}
\end{center}
\caption{\footnotesize{Leading-order diagrams in the scalar diquark model for the matrix element in Eq.~\eqref{eq:scalarDiquarkMEa}
(left panel) and in Eq.~\eqref{eq:scalarDiquarkMEb} (right panel). 
\label{fig:Leading-order-diagram}}}
\vspace{-0.2 truecm}
\end{figure}

We calculate the matrix element in Eq.~\eqref{qGTMD} in the scalar diquark model, i.e., a Yukawa theory defined by the Lagrangian (see also App.~A in Ref.~\cite{Meissner:2007rx})
\begin{equation}
\uL=\overline{\Psi}_{N}(i\uslash\partial-M)\Psi_{N}+\overline{\psi}(i\uslash\partial-m_{q})\psi+(\partial_{\mu}\phi\,\partial^{\mu}\phi^*-m_{s}^{2}|\phi|^{2}) + g_{s}(\overline{\psi}\Psi_{N}\phi^* + \overline{\Psi}_{N}\psi\phi),
\label{eq:diquarkLagrangian-1}
\end{equation}
where $\Psi_N$ denotes the fermionic target field with mass $M$, $\psi$ the quark field with mass $m_q$,  and $\phi$ the scalar diquark field with mass $m_s$.
To leading order in the coupling $g_{s}$ of the Yukawa interaction, the time-ordered quark correlator reads
\begin{align}
\langle p',\Lambda'|\mathrm{T}\{\overline{\psi}_{\alpha}(-\tfrac{z}{2})\psi_{\beta}(\tfrac{z}{2})\}|p,\Lambda\rangle&= ig_{s}^{2}\int\frac{\ud^4q}{(2\pi)^4}\,\frac{\left[\overline{u}'(\uslash q-m_{q}-M)\right]_{\alpha}\left[(\uslash q-m_{q}-M)u\right]_{\beta}}{D_q}\,e^{-i(P-q)\cdot z}\nonumber \\
&\quad+\mathcal O(g^4_s),\label{eq:scalarDiquarkMEa}
\end{align}
where the denominator is $D_q=\left[q^{2}-m_{s}^{2}+i\epsilon\right]\left[(p'-q)^{2}-m_{q}^{2}+i\epsilon\right]\left[(p-q)^{2}-m_{q}^{2}+i\epsilon\right]$, $\overline u'=\overline u(p',\Lambda')$ and $u=u(p,\Lambda)$. 
This expression can be depicted by the diagram on the left panel of Fig.~\ref{fig:Leading-order-diagram}. 
The diagram on the right panel of Fig.~\ref{fig:Leading-order-diagram} represents the scalar diquark contribution corresponding to the non-local correlator
\begin{equation}
\langle p',\Lambda'|\mathrm{T}\{\phi^*(-\tfrac{z}{2})\phi(\tfrac{z}{2})\}|p,\Lambda\rangle = -ig_{s}^{2}\int\frac{\ud^4q}{(2\pi)^4}\,\frac{\overline{u}'(\uslash q+m_{q})u}{D_s}\,e^{-i(P-q)\cdot z} + \mathcal O(g^4_s),
\label{eq:scalarDiquarkMEb}
\end{equation}
where the denominator is $D_s=\left[q^{2}-m_{q}^{2}+i\epsilon\right]\left[(p'-q)^{2}-m_{s}^{2}+i\epsilon\right]\left[(p-q)^{2}-m_{s}^{2}+i\epsilon\right]$.

Let us start with the quark contribution to $F_{1,4}$ which can be extracted from the correlator
\begin{equation}
\frac{1}{2}\int\ud\bar k^-\!\int\frac{\ud^4z}{(2\pi)^4}\,e^{i \bar k\cdot z}\,\langle p',\Lambda'|\overline\psi(-\tfrac{z}{2})\gamma^+\psi(\tfrac{z}{2})|p,\Lambda\rangle.
\end{equation}
Contracting Eq.~\eqref{eq:scalarDiquarkMEa} with $\gamma^+$, performing the Fourier transform from $z$-space to $\bar k$-space, and integrating over $\bar k^-$, we find by comparison with the GTMD parametrization of Ref.~\cite{Meissner:2009ww}:
\begin{equation}\label{SDMa}
F^q_{1,4}=-\frac{g^2_s}{2(2\pi)^3}\,\frac{(1-x)^2M^2}{\left[\uvec k'^2_\perp+\mathcal M^2(x)\right]\left[\uvec k^2_\perp+\mathcal M^2(x)\right]}+\mathcal O(g^4_s),
\end{equation}
where 
\begin{align}
\uvec k'_\perp&=\bar{\uvec k}_\perp+(1-x)\,\tfrac{\uvec \Delta_\perp}{2},\\
\uvec k_\perp&=\bar{\uvec k}_\perp-(1-x)\,\tfrac{\uvec \Delta_\perp}{2},\\
\mathcal M^2(x)&=(1-x)m^2_q+xm^2_s-x(1-x)M^2.
\end{align}
A similar calculation from the contraction of  the correlator \eqref{eq:scalarDiquarkMEa} with $\gamma^+\gamma_5$  leads to
\begin{equation}\label{SDMc}
G^q_{1,1}=-\frac{g^2_s}{2(2\pi)^3}\,\frac{(1-x)^2M^2}{\left[\uvec k'^2_\perp+\mathcal M^2(x)\right]\left[\uvec k^2_\perp+\mathcal M^2(x)\right]}+\mathcal O(g^4_s).
\end{equation}
So, to leading order in the scalar diquark model one finds $F^q_{1,4}=G_{1,1}^q$.
The expressions in Eqs.~(\ref{SDMa}), (\ref{SDMc}) are in full agreement with the (nonzero) results presented in Ref.~\cite{Meissner:2009ww}.

One may also study GTMDs with the scalar diquark acting as parton, which was not yet done in~\cite{Meissner:2009ww}.
Clearly, the scalar diquark cannot contribute to $G_{1,1}$ since this GTMD requires the active parton to be polarized. 
The scalar diquark contribution to $F_{1,4}$ can be extracted from the correlator
\begin{equation}
\frac{1}{2}\int\ud\bar k^-\!\int\frac{\ud^4z}{(2\pi)^4}\,e^{i \bar k\cdot z}\,\langle p',\Lambda'|\phi^*(-\tfrac{z}{2})i\overset{\leftrightarrow}{\partial}\phantom{\partial\!\!\!\!}^+\!\phi(\tfrac{z}{2})|p,\Lambda\rangle,
\end{equation}
where $\phi^*(-\tfrac{z}{2})i\overset{\leftrightarrow}{\partial}\phantom{\partial\!\!\!\!}^+\phi(\tfrac{z}{2})=\phi^*(-\tfrac{z}{2})[i\partial^+\!\phi(\tfrac{z}{2})]-[i\partial^+\!\phi(-\tfrac{z}{2})]\phi(\tfrac{z}{2})=2i\partial^+_z[\phi^*(-\tfrac{z}{2})\phi(\tfrac{z}{2})]$. As a result, we obtain
\begin{align}
F^s_{1,4}&=-\frac{g^2_s}{2(2\pi)^3}\,\frac{x(1-x)M^2}{\left[\uvec k'^2_\perp+\mathcal M^2(1-x)\right]\left[\uvec k^2_\perp+\mathcal M^2(1-x)\right]}+\mathcal O(g^4_s),\label{SDMb}\\
G^s_{1,1}&=0.
\end{align}

\subsection{Quark-Target Model}
\label{subs:QTM}
\begin{figure}[t!]
\begin{center}
\epsfig{file=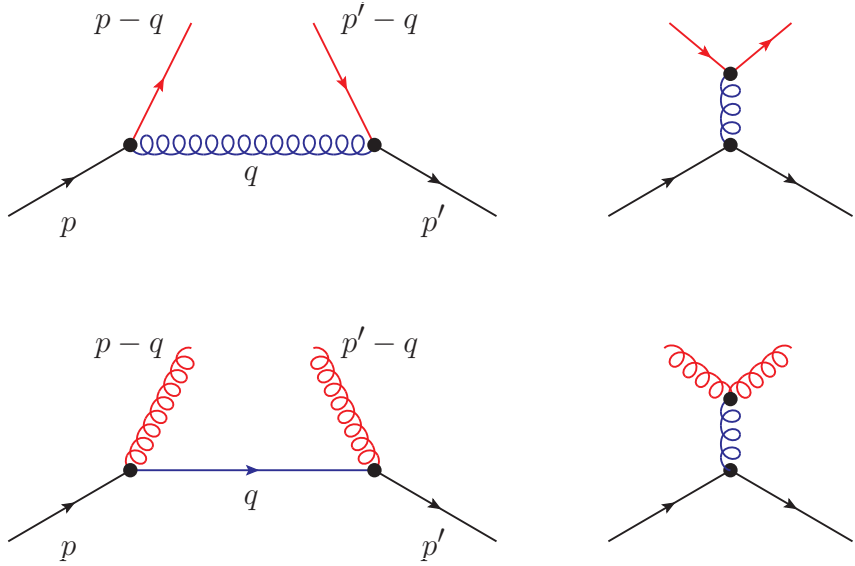,  width=.70\textwidth}
\end{center}
\caption{\footnotesize{Leading-order diagrams in the quark-target model (in light-front gauge) for the matrix element in Eq.~\eqref{eq:quarktargetMEa} (upper row) and in Eq.~\eqref{eq:quarktargetMEb} (lower row).
Note that for the amplitude in~(\ref{eq:quarktargetMEb}) we only consider diagrams which eventually could contribute to gluon GTMDs for $x\in[0,1]$.  
\label{fig:Leading-non-trivial-contribution}}}
\vspace{-0.2 truecm}
\end{figure}
We continue with the same strategy as above and calculate the GTMDs in QCD for a quark target $|p,\Lambda;a_{i}\rangle$, where $a_{i(f)}$ is the color of the initial (final) quark.
Then, a first non-trivial perturbative expression for the GTMDs can be extracted from the diagrams in Fig.~\ref{fig:Leading-non-trivial-contribution}.
We point out that, to the order in perturbation theory considered here, virtual radiative corrections do not contribute to $F_{1,4}$ and $G_{1,1}$.
For the quark contribution, we evaluate the following matrix element in perturbative QCD,
\begin{align}
\langle p',\Lambda';a_f|\mathrm{T}\{\overline{\psi}_{\alpha}(-\tfrac{z}{2})\mathcal W_{n_-}\psi_{\beta}(\tfrac{z}{2})\}|p,\Lambda;a_i\rangle&\nonumber \\
&\hspace{-6.5cm}=-4\pi i\,C_{F}\alpha_S\delta_{a_{i}a_{f}}\int\frac{\ud^4q}{(2\pi)^4}\,\frac{d_{\mu\nu}(q,n_-)\left[\overline{u}^{\prime}\gamma^{\mu}(\uslash p'-\uslash q+m_{q})\right]_{\alpha}\left[(\uslash p-\uslash q+m_{q})\gamma^{\nu}u\right]_{\beta}}{D_q}\,e^{-i(P-q)\cdot z}\nonumber\\
&\hspace{-6.5cm}+t\text{-channel}+\mathcal{O}(\alpha_S^{2}),\label{eq:quarktargetMEa}
\end{align}
where the denominator is $D_q=\left[q^{2}+i\epsilon\right]\left[(p'-q)^{2}-m_{q}^{2}+i\epsilon\right]\left[(p-q)^{2}-m_{q}^{2}+i\epsilon\right]$, and $C_F = (N_c^2 - 1)/2N_c = 4/3$. 
We work in the light-front gauge $A^+=0$, and so the numerator of the gluon propagator reads
\begin{equation}
d^{\mu\nu}(q,n_-)=g^{\mu\nu}-\frac{q^{\mu}n^{\nu}_-+q^{\nu}n^{\mu}_-}{q\cdot n_-}.
\label{eq:polsum}
\end{equation} 
For the gluon contribution, the correlator in perturbative QCD takes the form
\begin{align}
\langle p',\Lambda';a_f|\mathrm{T}\{A_{\perp\rho}(-\tfrac{z}{2})\mathcal W_{n_-} A_{\perp\sigma}(\tfrac{z}{2})\mathcal W_{n_-}\}|p,\Lambda;a_i\rangle&\nonumber \\
&\hspace{-7cm}=4\pi i\,C_{F}\alpha_S\delta_{a_{i}a_{f}}\int\frac{\ud^4q}{(2\pi)^4}\,\frac{d_{\mu\rho}(p^{\prime}-q,n_-)\,d_{\nu\sigma}(p-q,n_-)\left[\overline u'\gamma^{\mu}(\uslash q+m_{q})\gamma^{\nu}u\right]}{D_g}\,e^{-i(P-q)\cdot z}\nonumber\\
&\hspace{-7cm}+t\text{-channel}+\mathcal{O}(\alpha_S^{2}),\label{eq:quarktargetMEb}
\end{align}
where the denominator is $D_g=\left[q^{2}-m_{q}^{2}+i\epsilon\right]\left[(p^{\prime}-q)^{2}+i\epsilon\right]\left[(p-q)^{2}+i\epsilon\right]$. 

When computing GTMDs only the $u$-channel graphs on the left in Fig.~\ref{fig:Leading-non-trivial-contribution} survive.
The $t$-channel graphs vanish at this step for actually two reasons: first due to kinematics, and second after proper color-averaging needed for the calculation of GTMDs.
Once again, we start with the quark contribution to $F_{1,4}$ which can be extracted from the correlator
\begin{equation}\label{QTMF14}
\frac{1}{2}\int\ud\bar k^-\!\int\frac{\ud^4z}{(2\pi)^4}\,e^{i \bar k\cdot z}\,\langle p',\Lambda';a_f|\overline\psi(-\tfrac{z}{2})\gamma^+\mathcal W_{n_-}\psi(\tfrac{z}{2})|p,\Lambda;a_i\rangle.
\end{equation}
By comparison with the GTMD parametrization of Ref.~\cite{Meissner:2009ww} we find
\begin{equation}\label{QTMa}
F^q_{1,4}=\frac{C_F\alpha_S}{2\pi^2}\,\frac{(1-x^2)\,m^2_q}{\left[\uvec k'^2_\perp+(1-x)^2m^2_q\right]\left[\uvec k^2_\perp+(1-x)^2m^2_q\right]}+\mathcal O(\alpha^2_S).
\end{equation}
Similarly, replacing $\gamma^+$ in Eq.~\eqref{QTMF14} by $\gamma^+\gamma_5$, we find for the quark contribution to $G_{1,1}$:
\begin{equation}\label{QTMc}
G^q_{1,1}=-\frac{C_F\alpha_S}{2\pi^2}\,\frac{(1-x^2)\,m^2_q}{\left[\uvec k'^2_\perp+(1-x)^2m^2_q\right]\left[\uvec k^2_\perp+(1-x)^2m^2_q\right]}+\mathcal O(\alpha^2_S).
\end{equation}
Therefore, to leading order in the quark-target model one finds $F^q_{1,4}=-G_{1,1}^q$\footnote{The results in Eqs.~\eqref{QTMa} and\eqref{QTMc} have been confirmed also in Ref.~\cite{Mukherjee:2014nya}.} .
The fact that one gets the opposite relation between $F^q_{1,4}$ and $G^q_{1,1}$ in the quark-target model compared to the scalar diquark model is easy to understand from a mathematical point of view. 
In the quark-target model there is an extra Dirac matrix to anticommute with $\gamma_5$ compared to the scalar diquark model, leading to an extra minus sign in the relation between $F^q_{1,4}$ and $G^q_{1,1}$.

Now, the gluon contribution to $F_{1,4}$ can be extracted from the correlator
\begin{equation}
\int\ud\bar k^-\!\int\frac{\ud^4z}{(2\pi)^4}\,e^{i \bar k\cdot z}\,\langle p',\Lambda';a_f|-2g^{\rho\sigma}_\perp\uTr[A_{\perp\rho}(-\tfrac{z}{2})\mathcal W_{n_-} A_{\perp\sigma}(\tfrac{z}{2})\mathcal W_{n_-}]|p,\Lambda;a_i\rangle.
\end{equation}
As a result, we obtain for $x\in[0,1]$
\begin{equation}\label{QTMb}
F^g_{1,4}=\frac{C_F\alpha_S}{2\pi^2}\,\frac{(1-x)(2-x)\,m^2_q}{\left[\uvec k'^2_\perp+x^2m^2_q\right]\left[\uvec k^2_\perp+x^2m^2_q\right]}+\mathcal O(\alpha^2_S).
\end{equation}
Similarly, the gluon contribution to $G_{1,1}$ can be extracted from the correlator
\begin{equation}
\int\ud\bar k^-\!\int\frac{\ud^4z}{(2\pi)^4}\,e^{i \bar k\cdot z}\,\langle p',\Lambda';a_f|-2i\epsilon^{\rho\sigma}_\perp\uTr[A_{\perp\rho}(-\tfrac{z}{2})\mathcal W_{n_-} A_{\perp\sigma}(\tfrac{z}{2})\mathcal W_{n_-}]|p,\Lambda;a_i\rangle,
\end{equation}
where $\epsilon^{\rho\sigma}_\perp=\epsilon^{\rho\sigma n_+n_-}$ with $\epsilon_{0123}=+1$. This leads us to
\begin{equation}\label{QTMd}
G^g_{1,1}=-\frac{C_F\alpha_S}{2\pi^2}\,\frac{1-x}{x}\,\frac{\left[(1-x)^2+1\right]m^2_q}{\left[\uvec k'^2_\perp+x^2m^2_q\right]\left[\uvec k^2_\perp+x^2m^2_q\right]}+\mathcal O(\alpha^2_S).
\end{equation}

Obviously, in the quark-target model $F_{1,4}$ and $G_{1,1}$ are nonzero for both quarks and gluons.
We performed the calculation of those objects also in Feynman gauge and found the same results.
Note that in Feynman gauge, in the case of quark GTMDs, one also needs to take into account contributions due to the gauge link of the GTMD correlator (see also Ref.~\cite{Meissner:2007rx}).
While in general those terms matter for GTMDs, the explicit calculation shows that they do not contribute to $F_{1,4}$ and $G_{1,1}$ through $\mathcal{O}(\alpha_S)$, if one works with either future-pointing or past-pointing Wilson lines as we do.
This also demonstrates, in the context of a model calculation, that the real part of those two GTMDs does not depend on the direction of the gauge contour.

\subsection{Canonical Orbital Angular Momentum and Spin-Orbit Correlation}
\label{subs:OAM}
Following the same diagrammatic approach, we now calculate the canonical OAM. 
The corresponding operators for quarks, scalar diquarks and gluons with momentum fraction $x\in[0,1]$ are given in the light-front gauge by
\begin{align}
\hat{\mathcal{O}}^{q}_{\ell_z}(x,r^{-},\uvec r_\perp) & \equiv \frac{1}{2}\int\frac{\ud z^-}{2\pi}\,e^{ixP^+z^-}\left[\overline{\psi}(r^{-}-\tfrac{z^-}{2},\uvec r_{\perp})\,\gamma^+(\uvec r_\perp\times i\uvec\partial_\perp)_z\,\psi(r^{-}+\tfrac{z^-}{2},\uvec r_{\perp})\right]+\text{h.c.}, \\
\hat{\mathcal{O}}^{s}_{\ell_z}(x,r^{-},\uvec r_\perp)&\equiv  \int\frac{\ud z^-}{2\pi}\,e^{ixP^+z^-}\left[\partial^+\!\phi^*(r^{-}-\tfrac{z^-}{2},\uvec r_{\perp})\,(\uvec r_\perp\times \uvec\partial_\perp)_z\,\phi(r^{-}+\tfrac{z^-}{2},\uvec r_{\perp})\right]+\text{h.c.}, \\
\hat{\mathcal{O}}^{g}_{\ell_z}(x,r^{-},\uvec r_\perp) & \equiv -\int\frac{\ud z^-}{2\pi}\,e^{ixP^+z^-}2g^{\rho\sigma}_{\perp}\,\Tr\!\left[\partial^+\!A_{\perp\rho}(r^{-}-\tfrac{z^-}{2},\uvec r_{\perp})\,(\uvec r_\perp\times \uvec\partial_\perp)_z\, A_{\perp\sigma}(r^{-}+\tfrac{z^-}{2},\uvec r_{\perp})\right]
\nonumber \\
& \hspace{0.5cm} +\text{h.c.}.\label{eq:OAMoperators}
\end{align}
We calculate also the canonical spin-orbit correlation~\cite{Lorce:2011kd,Lorce:2014mxa}, where the corresponding operators for quarks and gluons are given by\footnote{As already mentioned in Sect.~\ref{subs:SDM}, scalar diquarks obviously cannot have a spin-orbit correlation.} 
\begin{align}
\hat{\mathcal{O}}^{q}_{C_z}(x,r^{-},\uvec r_\perp) & \equiv \frac{1}{2}\int\frac{\ud z^-}{2\pi}\,e^{ixP^+z^-}\left[\overline{\psi}(r^{-}-\tfrac{z^-}{2},\uvec r_{\perp})\,\gamma^+\gamma_5(\uvec r_\perp\times i\uvec\partial_\perp)_z\,\psi(r^{-}+\tfrac{z^-}{2},\uvec r_{\perp})\right]+\text{h.c.}, \\
\hat{\mathcal{O}}^{g}_{C_z}(x,r^{-},\uvec r_\perp) & \equiv -\int\frac{\ud z^-}{2\pi}\,e^{ixP^+z^-}2i\epsilon^{\rho\sigma}_{\perp}\,\Tr\!\left[\partial^+\!A_{\perp\rho}(r^{-}-\tfrac{z^-}{2},\uvec r_{\perp})\,(\uvec r_\perp\times \uvec\partial_\perp)_z\, A_{\perp\sigma}(r^{-}+\tfrac{z^-}{2},\uvec r_{\perp})\right]
\nonumber \\
& \hspace{0.5cm} +\text{h.c.}.\label{eq:SOCoperators}
\end{align}
Because of the explicit factor $\uvec r_\perp$, one has to be careful when considering the expectation value of these operators~\cite{Hagler:2003jw,Bakker:2004ib,Leader:2013jra}:
\begin{align}
\ell^{q,s,g}_z(x)&\equiv\lim_{\Delta\to 0}\frac{P^+ \int\ud r^{-} \int\ud^2{\uvec r}_\perp\,\langle  p', \Lambda |  \hat{\mathcal{O}}^{q,s,g}_{\ell_z}(x,r^{-},\uvec r_\perp) | p, \Lambda  \rangle}{\langle p',\Lambda|p,\Lambda\rangle},\\
C^{q,g}_z(x)&\equiv\lim_{\Delta\to 0}\frac{P^+ \int\ud r^{-} \int\ud^2{\uvec r}_\perp\,\langle  p', \Lambda |  \hat{\mathcal{O}}^{q,g}_{C_z}(x,r^{-},\uvec r_\perp) | p, \Lambda  \rangle}{\langle p',\Lambda|p,\Lambda\rangle}.
\end{align}

Using the leading-order expressions \eqref{eq:scalarDiquarkMEa} and \eqref{eq:scalarDiquarkMEb} for the correlators in the scalar diquark model we obtain
\begin{align} \label{eq:lq_SDM}
\ell^q_z(x)&=\frac{g^2_s}{2(2\pi)^3}\int\ud^2\bar{\uvec k}_\perp\,\frac{(1-x)^2\,\bar{\uvec k}^2_\perp}{\left[\bar{\uvec k}^2_\perp+\mathcal M^2(x)\right]^2}+\mathcal O(g^4_s),\\
\ell^s_z(x)&=\frac{g^2_s}{2(2\pi)^3}\int\ud^2\bar {\uvec k}_\perp\,\frac{x(1-x)\,\bar{\uvec k}^2_\perp}{\left[\bar{\uvec k}^2_\perp+\mathcal M^2(1-x)\right]^2}+\mathcal O(g^4_s),\\
\label{eq:Cq_SDM}
C^q_z(x)&=-\frac{g^2_s}{2(2\pi)^3}\int\ud^2\bar {\uvec k}_\perp\,\frac{(1-x)^2\,\bar{\uvec k}^2_\perp}{\left[\bar{\uvec k}^2_\perp+\mathcal M^2(x)\right]^2}+\mathcal O(g^4_s).
\end{align}
We note that the result for $\ell^q_z$ in Eq.~(\ref{eq:lq_SDM}) can already be found in Ref.~\cite{Burkardt:2008ua}.
Comparing the expressions in~(\ref{eq:lq_SDM})--(\ref{eq:Cq_SDM}) with Eqs.~\eqref{SDMa}, \eqref{SDMc} and \eqref{SDMb} we see that
\begin{align}
\ell^{q,s}_z(x)&=-\int\ud^2\bar {\uvec k}_\perp\,\frac{\bar{\uvec k}^2_\perp}{M^2}\,F^{q,s}_{1,4}(x,0,\bar{ \uvec k}^2_\perp,0,0),\label{relA}\\
C^{q}_z(x)&=\int\ud^2\bar {\uvec k}_\perp\,\frac{\bar{\uvec k}^2_\perp}{M^2}\,G^{q}_{1,1}(x,0,\bar{ \uvec k}^2_\perp,0,0),\label{relB}
\end{align}
which is an explicit check of the model-independent relations\footnote{Note that in Refs.~\cite{Lorce:2011kd,Lorce:2011ni,Hatta:2011ku,Lorce:2014mxa} the relation between the canonical OAM and the GTMD $F_{1,4}$ is integrated over $x$ for convenience. Since this integration does not affect the derivation, the relation remains valid at the $x$-density level.} of Refs.~\cite{Lorce:2011kd,Lorce:2011ni,Hatta:2011ku,Lorce:2014mxa}. 
Since in the scalar diquark model $F^q_{1,4}=G^q_{1,1}$, we have the result $\ell^q_z=-C^q_z$. 
If we now denote the quark momentum fraction by $x$, the total OAM simply reads
\begin{align}
\ell_z(x)&=\ell^q_z(x)+\ell^s_z(1-x)\nonumber\\
&=\frac{g^2_s}{2(2\pi)^3}\int\ud^2\bar {\uvec k}_\perp\,\frac{(1-x)\,\bar{\uvec k}^2_\perp}{\left[\bar{\uvec k}^2_\perp+\mathcal M^2(x)\right]^2}+\mathcal O(g^4_s),
\end{align}
and we have
\begin{align}
\ell^q_z(x)&=(1-x)\,\ell_z(x),\\
\ell^s_z(1-x)&=x\,\ell_z(x),
\end{align}
in agreement with the discussion in Refs.~\cite{Harindranath:1998ve,Burkardt:2008ua}. 
Of course such simple relations between partial and total OAM hold only for a two-body system.

Similarly, using the leading-order expressions \eqref{eq:quarktargetMEa} and \eqref{eq:quarktargetMEb} for the correlators in the quark-target model, we obtain
\begin{align} \label{eq:lq_QTM}
\ell^q_z(x)&=-\frac{C_F\alpha_S}{2\pi^2}\int\ud^2\bar {\uvec k}_\perp\,\frac{(1-x^2)\,\bar{\uvec k}^2_\perp}{\left[\bar{\uvec k}^2_\perp+(1-x)^2m^2_q\right]^2}+\mathcal O(\alpha^2_S),\\
\label{eq:lg_QTM}
\ell^g_z(x)&=-\frac{C_F\alpha_S}{2\pi^2}\int\ud^2\bar {\uvec k}_\perp\,\frac{(1-x)(2-x)\,\bar{\uvec k}^2_\perp}{\left[\bar{\uvec k}^2_\perp+x^2m^2_q\right]^2}+\mathcal O(\alpha^2_S),\\
C^q_z(x)&=-\frac{C_F\alpha_S}{2\pi^2}\int\ud^2\bar {\uvec k}_\perp\,\frac{(1-x^2)\,\bar{\uvec k}^2_\perp}{\left[\bar{\uvec k}^2_\perp+(1-x)^2m^2_q\right]^2}+\mathcal O(\alpha^2_S),\\
\label{eq:Cg_QTM}
C^g_z(x)&=-\frac{C_F\alpha_S}{2\pi^2}\int\ud^2\bar {\uvec k}_\perp\,\frac{1-x}{x}\,\frac{[(1-x)^2+1]\,\bar{\uvec k}^2_\perp}{\left[\bar{\uvec k}^2_\perp+x^2m^2_q\right]^2}+\mathcal O(\alpha^2_S).
\end{align}
In the quark-target model, the canonical OAM in Eqs.~(\ref{eq:lq_QTM}), (\ref{eq:lg_QTM}) for quarks and gluons was studied for the first time in Ref.~\cite{Harindranath:1998ve} with a focus on the ultraviolet-divergent part (see also Ref.~\cite{Burkardt:2008ua}).
Comparing the expressions in~(\ref{eq:lq_QTM})--(\ref{eq:Cg_QTM}) with Eqs.~\eqref{QTMa}, \eqref{QTMc}, \eqref{QTMb} and \eqref{QTMd}, we see that
\begin{align}
\ell^{q,g}_z(x)&=-\int\ud^2\bar {\uvec k}_\perp\,\frac{\bar{\uvec k}^2_\perp}{m^2_q}\,F^{q,g}_{1,4}(x,0,\bar{ \uvec k}^2_\perp,0,0),\label{relC}\\
C^{q,g}_z(x)&=\int\ud^2\bar {\uvec k}_\perp\,\frac{\bar{\uvec k}^2_\perp}{m^2_q}\,G^{q,g}_{1,1}(x,0,\bar{ \uvec k}^2_\perp,0,0),\label{relD}
\end{align}
which is another explicit check of the model-independent relations of Refs.~\cite{Lorce:2011kd,Lorce:2011ni,Hatta:2011ku,Lorce:2014mxa}. Since in the quark-target model $F^q_{1,4}=-G^q_{1,1}$, we find now $\ell^q_z=C^q_z$. 
If we again denote the quark momentum fraction by $x$, the total OAM reads
\begin{align}
\ell_z(x)&=\ell^q_z(x)+\ell^g_z(1-x)\nonumber\\
&=-\frac{C_F\alpha_S}{2\pi^2}\int\ud^2\bar {\uvec k}_\perp\,\frac{(1+x)\,\bar{\uvec k}^2_\perp}{\left[\bar{\uvec k}^2_\perp+(1-x)^2m^2_q\right]^2}+\mathcal O(\alpha^2_S),
\end{align}
and we have
\begin{align}
\ell^q_z(x)&=(1-x)\,\ell_z(x),\\
\ell^g_z(1-x)&=x\,\ell_z(x).
\end{align}
We note in passing that the quark OAM defined through $F_{1,4}$ is the same for a future-pointing and for a past-pointing gauge contour of the GTMDs (see the paragraph after Eq.~(\ref{QTMd})).
This result was already put forward in Ref.~\cite{Hatta:2011ku}, and, to the best of our knowledge, we have now verified it for the first time explicitly in a model calculation.

%
%
\section{Overlap representation}
\label{s:Overlap}
The results obtained in the previous Sect.~\ref{s:Models} using a diagrammatic approach within the scalar diquark model and the quark-target model can also be derived in a straightforward way by using the overlap representation in terms of LFWFs.
As the LFWFs are eigenstates of the canonical OAM operator, in this approach the results for the OAM and the spin-orbit correlations have a transparent and intuitive interpretation.

\subsection{Scalar Diquark Model}
\label{subs:SDM2}
In the scalar diquark model, to lowest non-trivial order in perturbation theory, the quark LFWFs $\psi^\Lambda_\lambda$ read
\begin{align}
\psi^{\uparrow}_{\uparrow}(x, \uvec k_\perp)&=\frac{m_q+xM}{x}\,\phi(x,\uvec k_\perp^2),\label{WF1}\\
\psi^{\uparrow}_{\downarrow}(x, \uvec k_\perp)&=-\frac{k^R}{x}\,\phi(x,\uvec k_\perp^2),\label{WF2}
\end{align}
with $\{\uparrow,\downarrow\}=\{+\tfrac{1}{2},-\tfrac{1}{2}\}$, and
\begin{equation}
\phi(x,\uvec k^2_\perp)=\frac{g_s}{\sqrt{1-x}}\,\frac{1}{M^2-\frac{\uvec k^2_\perp+m_q^2}{x}-\frac{\uvec k^2_\perp+m_s^2}{1-x}}=-\frac{g_s}{\sqrt{1-x}}\,\frac{x(1-x)}{\uvec k^2_\perp+\mathcal M^2(x)}.
\end{equation}
The LFWFs with opposite nucleon helicity follow directly from light-front parity and time-reversal symmetries according to
\begin{equation}
\psi^{\Lambda}_{\lambda}(x, \uvec k_\perp)=\psi^{-\Lambda}_{-\lambda}(x, \uvec k_{\perp\mathsf P})=(-1)^{l_z}\,\psi^{-\Lambda\ast}_{-\lambda}(x, \uvec k_\perp),
\end{equation}
with $l_z=\Lambda-\lambda$ the total OAM associated with the LFWF $\psi^\Lambda_\lambda$. 
The corresponding scalar diquark LFWFs are simply obtained by replacing the argument $(x,\uvec k_\perp)$ of the quark LFWFs by $(\hat x,\hat{\uvec k}_\perp)=(1-x,-\uvec k_\perp)$, owing to momentum conservation. 

The LFWF overlap representation of the quark contribution to $F_{1,4}$ and $G_{1,1}$ at $\xi=0$ is then given by
\begin{align}
\tfrac{i(\bar{\uvec k}_\perp\times\uvec\Delta_\perp)_z}{M^2}\,F_{1,4}^q &=\frac{1}{2(2\pi)^3}\,\frac{1}{2}\sum_{\Lambda,\lambda}\text{sign}(\Lambda)\left[ 
\psi^{\Lambda\ast}_{\lambda}(x,\uvec k'_\perp) \psi^{\Lambda}_{\lambda}(x,\uvec k_\perp)
\right]\nonumber\\
&=\frac{i}{2(2\pi)^3}\,\Im\!\left[\psi^{\uparrow\ast}_{\downarrow}(x,\uvec k'_\perp) \psi^{\uparrow}_{\downarrow}(x,\uvec k_\perp)\right],\label{F14:LFWF}\\
-\tfrac{i(\bar{\uvec k}_\perp\times\uvec\Delta_\perp)_z}{M^2}\,G_{1,1}^q&=\frac{1}{2(2\pi)^3}\,\frac{1}{2}\sum_{\Lambda,\lambda}\text{sign}(\lambda)
\left[ \psi^{\Lambda\ast}_{\lambda}(x,\uvec k'_\perp) \psi^{\Lambda}_{\lambda}(x,\uvec k_\perp)
\right]
\nonumber\\
&=-\frac{i}{2(2\pi)^3}\,\Im\!\left[\psi^{\uparrow\ast}_{\downarrow}(x,\uvec k'_\perp) \psi^{\uparrow}_{\downarrow}(x,\uvec k_\perp)\right].
\label{G11:LFWF}
\end{align}
From Eqs.~\eqref{F14:LFWF} and \eqref{G11:LFWF}, we immediately see that $F^q_{1,4}=G_{1,1}^q$ as already observed in Sect.~\ref{subs:SDM}. 
For the scalar diquark contributions, we obviously have
\begin{align}
\tfrac{i(\bar{\uvec k}_\perp\times\uvec\Delta_\perp)_z}{M^2}\,F_{1,4}^s &=\frac{1}{2(2\pi)^3}\,\frac{1}{2}\sum_{\Lambda,\lambda}\text{sign}(\Lambda)\left[ 
\psi^{\Lambda\ast}_{\lambda}(\hat x,\hat{\uvec k}\phantom{k\!\!\!}'_\perp) \psi^{\Lambda}_{\lambda}(\hat x,\hat{\uvec k}_\perp)
\right]\nonumber\\
&=\frac{i}{2(2\pi)^3}\,\Im\!\left[\psi^{\uparrow\ast}_{\downarrow}(\hat x,\hat{\uvec k}\phantom{k\!\!\!}'_\perp) \psi^{\uparrow}_{\downarrow}(\hat x,\hat{\uvec k}_\perp)\right],\\
-\tfrac{i(\bar{\uvec k}_\perp\times\uvec\Delta_\perp)_z}{M^2}\,G_{1,1}^s&=0.
\end{align}
Using the explicit expressions~\eqref{WF1}-\eqref{WF2} for the quark LFWFs in the scalar diquark model, we reproduce the results of Sect.~\ref{subs:SDM}.

Since the ``parton'' OAM is simply given by $(1-x)$ times the total OAM in a two-body system~\cite{Harindranath:1998ve,Burkardt:2008ua}, we can easily write down the overlap representation of the canonical OAM and spin-orbit correlation for the quark,
\begin{align}
\ell^q_z(x)&=(1-x)\int\frac{\ud^2\bar {\uvec k}_\perp}{2(2\pi)^3}\,\frac{1}{2}\sum_{\Lambda,\lambda}\text{sign}(\Lambda)\,l_z\,|\psi^\Lambda_\lambda(x,\bar{\uvec k}_\perp)|^2\nonumber\\
&=(1-x)\int\frac{\ud^2\bar {\uvec k}_\perp}{2(2\pi)^3}\,|\psi^\uparrow_\downarrow(x,\bar{\uvec k}_\perp)|^2,\\
C^q_z(x)&=(1-x)\int\frac{\ud^2\bar {\uvec k}_\perp}{2(2\pi)^3}\,\frac{1}{2}\sum_{\Lambda,\lambda}\text{sign}(\lambda)\,l_z\,|\psi^\Lambda_\lambda(x,\bar{\uvec k}_\perp)|^2\nonumber\\
&=-(1-x)\int\frac{\ud^2\bar {\uvec k}_\perp}{2(2\pi)^3}\,|\psi^\uparrow_\downarrow(x,\bar{\uvec k}_\perp)|^2,
\end{align}
and the scalar diquark,
\begin{align}
\ell^s_z(x)&=(1-x)\int\frac{\ud^2\hat{\bar {\uvec k}}_\perp}{2(2\pi)^3}\,\frac{1}{2}\sum_{\Lambda,\lambda}\text{sign}(\Lambda)\,l_z\,|\psi^\Lambda_\lambda(\hat x,\hat{\bar{\uvec k}}_\perp)|^2\nonumber\\
&=(1-x)\int\frac{\ud^2\bar {\uvec k}_\perp}{2(2\pi)^3}\,|\psi^\uparrow_\downarrow(1-x,\bar{\uvec k}_\perp)|^2,\\
C^s_z(x)&=0.
\end{align}
The relation $\ell^q_z=-C^q_z$ can now be readily understood in physical terms.
In the scalar diquark model, the total OAM associated with a LFWF $l_z=\Lambda-\lambda$ is necessarily correlated with the nucleon helicity $\Lambda$ and anti-correlated with the quark helicity $\lambda$ by angular momentum conservation. 
The ``parton'' OAM being simply $(1-x)$ times the total OAM, we have automatically $\ell^q_z=-C^q_z>0$ and $\ell^s_z>0$. 
Finally, we have
\begin{equation}
\Im\!\left[\psi^{\uparrow\ast}_{\downarrow}(x,\uvec k'_\perp) \psi^{\uparrow}_{\downarrow}(x,\uvec k_\perp)\right]=-(1-x)\,\frac{(\bar{\uvec k}_\perp\times\uvec\Delta_\perp)_z}{\bar{\uvec k}^2_\perp}\,|\psi^\uparrow_\downarrow(x,\bar{\uvec k}_\perp)|^2+\mathcal O(\uvec\Delta^2_\perp),
\end{equation}
from which follow Eqs.~\eqref{relA} and \eqref{relB}.

\subsection{Quark-Target Model}
\label{subs:QTM2}
In the quark-target model, the quark LFWF is given by~\cite{Harindranath:1998ve,Brodsky:2000ii,Harindranath:2001rc,Kundu:2001pk} 
\begin{equation}
\psi^\Lambda_{\lambda,\mu}(x,\uvec k_\perp)=\Phi(x,\uvec k_\perp^2)\,\chi^\dagger_{\lambda}\left[- 2\,\tfrac{\uvec k_\perp}{1-x}-\tfrac{\tilde{\uvec\sigma}_\perp\cdot \uvec k_\perp}{x}\,\tilde{\uvec \sigma}_\perp
+i m_q\,{\tilde{\uvec\sigma}}_\perp\,\tfrac{1-x}{x}\right]\cdot
\uvec{\epsilon}_{\perp}^*(\mu)\chi_\Lambda,
\label{psi2}
\end{equation}
where $\tilde\sigma_1=\sigma_2$ and $\tilde \sigma_2=-\sigma_1$, the polarization vectors are defined as $\uvec{\epsilon}_\perp(\mu=+1)=-\tfrac{1}{\sqrt{2}}(1,i)$ and $\uvec{\epsilon}_\perp(\mu=-1)=\tfrac{1}{\sqrt{2}}(1,-i)$, and
\begin{equation}
\Phi(x,\uvec k_\perp^2)=\frac{gT}{\sqrt{1-x}}\,\frac{x(1-x)}{\uvec k^2_\perp+(1-x)^2m_q^2},
\end{equation}
with $T$ a color-$SU(3)$ generator. 
Specifying the helicity state in Eq.~\eqref{psi2}, one obtains
\begin{align}
\psi^{\uparrow}_{\uparrow,\Uparrow}(x, \uvec k_\perp)&=\frac{\sqrt{2}\,k^L}{x(1-x)}\,\Phi(x,\uvec k_\perp^2),\label{TWF1}\\
\psi^{\uparrow}_{\downarrow,\Uparrow}(x, \uvec{k}_\perp)&=\frac{\sqrt{2}\,m_q(1-x)}{x}\,\Phi(x,\uvec k_\perp^2),\label{TWF2}\\
\psi^{\uparrow}_{\uparrow,\Downarrow}(x, \uvec k_\perp)&=-\frac{\sqrt{2}\,k^R}{1-x}\,\Phi(x,\uvec k_\perp^2),\label{TWF3}\\
\psi^{\uparrow}_{\downarrow,\Downarrow}(x, \uvec k_\perp)&=0,\label{TWF4}
\end{align}
with $\{\Uparrow,\Downarrow\}=\{+1,-1\}$. 
Once again, the LFWFs with opposite nucleon helicity follow directly from light-front parity and time-reversal\footnote{Note that the LFWFs here, in principle, could also depend on $\eta$ due to initial/final-state interactions. However, as already discussed in Sect.~\ref{subs:QTM}, such potential effects are not relevant for the present study.} symmetries, namely
\begin{equation}
\psi^{\Lambda}_{\lambda,\mu}(x, \uvec k_\perp)=\psi^{-\Lambda}_{-\lambda,-\mu}(x, \uvec k_{\perp\mathsf P})=(-1)^{l_z}\,\psi^{-\Lambda\ast}_{-\lambda,-\mu}(x, \uvec k_\perp),
\end{equation}
with $l_z=\Lambda-\lambda-\mu$ the total OAM associated with the LFWF $\psi^\Lambda_{\lambda,\mu}$. 
The corresponding gluon LFWFs are simply obtained by replacing the argument $(x,\uvec k_\perp)$ of the quark LFWFs by $(\hat x,\hat{\uvec k}_\perp)=(1-x,-\uvec k_\perp)$, owing to momentum conservation. 

The LFWF overlap representation of the quark contribution to $F_{1,4}$ and $G_{1,1}$ is
\begin{align}
\tfrac{i(\bar{\uvec k}_\perp\times\uvec\Delta_\perp)_z}{m^2_q}\,F_{1,4}^q &=\frac{1}{2(2\pi)^3}\,\frac{1}{2}\sum_{\Lambda,\lambda,\mu}\text{sign}(\Lambda)\left[ 
\psi^{\Lambda\ast}_{\lambda,\mu}(x,\uvec k'_\perp) \psi^{\Lambda}_{\lambda,\mu}(x,\uvec k_\perp)
\right]\nonumber\\
&=\frac{i}{2(2\pi)^3}\sum_\mu\Im\!\left[\psi^{\uparrow\ast}_{\uparrow,\mu}(x,\uvec k'_\perp) \psi^{\uparrow}_{\uparrow,\mu}(x,\uvec k_\perp)\right],\label{F14:LFWF2}\\
-\tfrac{i(\bar{\uvec k}_\perp\times\uvec\Delta_\perp)_z}{m^2_q}\,G_{1,1}^q &=\frac{1}{2(2\pi)^3}\,\frac{1}{2}\sum_{\Lambda,\lambda,\mu}\text{sign}(\lambda)\left[ 
\psi^{\Lambda\ast}_{\lambda,\mu}(x,\uvec k'_\perp) \psi^{\Lambda}_{\lambda,\mu}(x,\uvec k_\perp)
\right]\nonumber\\
&=\frac{i}{2(2\pi)^3}\sum_\mu\Im\!\left[\psi^{\uparrow\ast}_{\uparrow,\mu}(x,\uvec k'_\perp) \psi^{\uparrow}_{\uparrow,\mu}(x,\uvec k_\perp)
\right].\label{G11:LFWF2}
\end{align}

From Eqs.~\eqref{F14:LFWF2} and \eqref{G11:LFWF2}, we immediately see that $F^q_{1,4}=-G_{1,1}^q$ in agreement with the observation in Sect.~\ref{subs:QTM}. 
Likewise, for the gluon contribution we have
\begin{align}
\tfrac{i(\bar{\uvec k}_\perp\times\uvec\Delta_\perp)_z}{m^2_q}\,F_{1,4}^g&=\frac{1}{2(2\pi)^3}\,\frac{1}{2}\sum_{\Lambda,\lambda,\mu}\text{sign}(\Lambda)\left[ 
\psi^{\Lambda\ast}_{\lambda,\mu}(\hat x,\hat{\uvec k}\phantom{k\!\!\!}'_\perp) \psi^{\Lambda}_{\lambda,\mu}(\hat x,\hat{\uvec k}_\perp)
\right]\nonumber\\
&=\frac{i}{2(2\pi)^3}\sum_\mu\Im\!\left[\psi^{\uparrow\ast}_{\uparrow,\mu}(\hat x,\hat{\uvec k}\phantom{k\!\!\!}'_\perp) \psi^{\uparrow}_{\uparrow,\mu}(\hat x,\hat{\uvec k}_\perp)\right],\\
-\tfrac{i(\bar{\uvec k}_\perp\times\uvec\Delta_\perp)_z}{m^2_q}\,G_{1,1}^g&=\frac{1}{2(2\pi)^3}\,\frac{1}{2}\sum_{\Lambda,\lambda,\mu}{\rm sign}(\mu)\left[ 
\psi^{\Lambda\ast}_{\lambda,\mu}(\hat x,\hat{\uvec k}\phantom{k\!\!\!}'_\perp) \psi^{\Lambda}_{\lambda,\mu}(\hat x,\hat{\uvec k}_\perp)
\right]\nonumber\\
&=\frac{i}{2(2\pi)^3}\,\sum_\mu{\rm sign}(\mu)\,\Im\!\left[\psi^{\uparrow\ast}_{\uparrow,\mu}(\hat x,\hat{\uvec k}\phantom{k\!\!\!}'_\perp) \psi^{\uparrow}_{\uparrow,\mu}(\hat x,\hat{\uvec k}_\perp)
\right].
\end{align}
Using the explicit expressions~\eqref{TWF1}-\eqref{TWF4} for the quark LFWFs in the quark-target model, we reproduce the results of Sect.~\ref{subs:QTM}.

We can also easily write down the overlap representation of the canonical OAM and spin-orbit correlation for the quark
\begin{align}
\ell^q_z(x)&=(1-x)\int\frac{\ud^2\bar {\uvec k}_\perp}{2(2\pi)^3}\,\frac{1}{2}\sum_{\Lambda,\lambda,\mu}\text{sign}(\Lambda)\,l_z\,|\psi^\Lambda_{\lambda,\mu}(x,\bar{\uvec k}_\perp)|^2\nonumber\\
&=-(1-x)\int\frac{\ud^2\bar {\uvec k}_\perp}{2(2\pi)^3}\sum_\mu\text{sign}(\mu)\,|\psi^\uparrow_{\uparrow,\mu}(x,\bar{\uvec k}_\perp)|^2,\\
C^q_z(x)&=(1-x)\int\frac{\ud^2\bar {\uvec k}_\perp}{2(2\pi)^3}\,\frac{1}{2}\sum_{\Lambda,\lambda,\mu}\text{sign}(\lambda)\,l_z\,|\psi^\Lambda_{\lambda,\mu}(x,\bar{\uvec k}_\perp)|^2\nonumber\\
&=-(1-x)\int\frac{\ud^2\bar {\uvec k}_\perp}{2(2\pi)^3}\sum_\mu\text{sign}(\mu)\,|\psi^\uparrow_{\uparrow,\mu}(x,\bar{\uvec k}_\perp)|^2,
\end{align}
and the gluon
\begin{align}
\ell^g_z(x)&=(1-x)\int\frac{\ud^2\hat{\bar {\uvec k}}_\perp}{2(2\pi)^3}\,\frac{1}{2}\sum_{\Lambda,\lambda,\mu}\text{sign}(\Lambda)\,l_z\,|\psi^\Lambda_{\lambda,\mu}(\hat x,\hat{\bar{\uvec k}}_\perp)|^2\nonumber\\
&=-(1-x)\int\frac{\ud^2\bar {\uvec k}_\perp}{2(2\pi)^3}\sum_\mu\text{sign}(\mu)\,|\psi^\uparrow_{\uparrow,\mu}(1-x,\bar{\uvec k}_\perp)|^2,\\
C^g_z(x)&=(1-x)\int\frac{\ud^2\hat{\bar {\uvec k}}_\perp}{2(2\pi)^3}\,\frac{1}{2}\sum_{\Lambda,\lambda,\mu}\text{sign}(\mu)\,l_z\,|\psi^\Lambda_{\lambda,\mu}(\hat x,\hat{\bar{\uvec k}}_\perp)|^2\nonumber\\
&=-(1-x)\int\frac{\ud^2\bar {\uvec k}_\perp}{2(2\pi)^3}\sum_\mu|\psi^\uparrow_{\uparrow,\mu}(1-x,\bar{\uvec k}_\perp)|^2.
\end{align}
The overlap representation provides an intuitive understanding of the relation $\ell^q_z=C^q_z$.
In the quark-target model, it turns out that only the LFWFs with correlated target and quark helicities contribute to the total OAM. 
The quark OAM being simply $(1-x)$ times the total OAM, we have automatically $\ell^q_z=C^q_z$. 
Moreover, for correlated target and quark helicities we have $l_z=-\mu$ which implies $C^g_z<0$. 
Finally, we have
\begin{equation}
\Im\!\left[\psi^{\uparrow\ast}_{\uparrow,\mu}(x,\uvec k'_\perp) \psi^{\uparrow}_{\uparrow,\mu}(x,\uvec k_\perp)\right]=(1-x)\,\frac{(\bar{\uvec k}_\perp\times\uvec\Delta_\perp)_z}{\bar{\uvec k}^2_\perp}\,\text{sign}(\mu)\,|\psi^\uparrow_{\uparrow,\mu}(x,\bar{\uvec k}_\perp)|^2+\mathcal O(\uvec\Delta^2_\perp),
\end{equation}
from which follow Eqs.~\eqref{relC} and \eqref{relD}.

%
%
\section{Perturbative tail of the GTMDs}
\label{s:PQCD}
\begin{figure}[t!]
\begin{center}
\epsfig{file=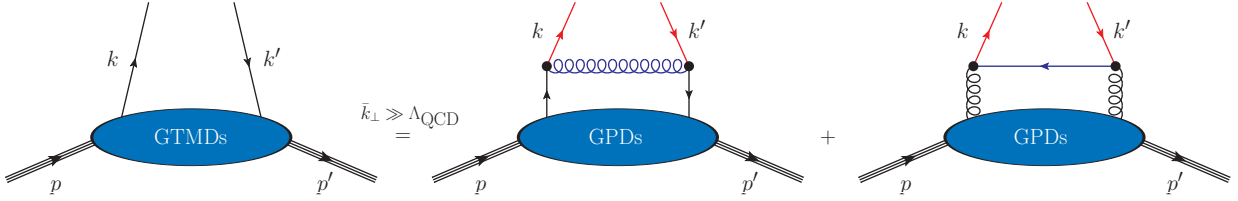,  width=1.0\textwidth}
\end{center}
\caption{\footnotesize{Diagrams determining the leading order of quark GTMDs at large transverse momenta.}
\label{fig:GTMDquark}}
\vspace{-0.2 truecm}
\end{figure}

In this section we present results for $F_{1,4}$ and $G_{1,1}$ at large transverse parton momenta, where they can be computed in perturbative QCD.
For simplicity we restrict the analysis to the real part of the GTMDs.
Details of a corresponding calculation of the large-$\bar k_\perp$ behavior of TMDs can be found in Ref.~\cite{Bacchetta:2008xw}.
Here we generalize these calculations to nonzero momentum transfer to the nucleon.
Like in Sect.~\ref{subs:QTM}, we have used both the light-front gauge $A^+ = 0$ and the Feynman gauge and have obtained the same results.

While for the quark we consider the correlator in Eq.~(\ref{qGPCF}), for the gluon we have
\begin{align}
W^{\mu\nu;\rho\sigma}_{\Lambda'\Lambda}(P,\bar k,\Delta,n_-)&=\frac{1}{\bar k^{+}}\int\frac{\ud^4z}{(2\pi)^4}\,e^{i \bar k\cdot z}\,\langle p',\Lambda'|T\{2\Tr[G^{\mu\nu}(-\tfrac{z}{2})\mathcal W_{n_-}G^{\rho\sigma}(\tfrac{z}{2})\mathcal W_{n_-}]\}|p,\Lambda\rangle.
\end{align}
For the calculation of $F_{1,4}$ and $G_{1,1}$ through $\mathcal{O}(\alpha_S)$ the light-like Wilson lines in those correlators do not give rise to the infamous light-cone singularities.
In fact, we have also performed a study with Wilson lines that are slightly off the light-cone, and in the end one can take the light-like limit without encountering a divergence.
This no longer necessarily holds for other GTMDs at large transverse momenta.

In the light-front gauge, the large-$\bar k_\perp$ behavior of the quark GTMDs is described perturbatively by the two diagrams on the right of Fig.~\ref{fig:GTMDquark}. 
Expressing them by means of Feynman rules leads to the following formula,
\begin{align}
&W_{\Lambda'\Lambda}^{[\Gamma]}(P,x,\bar{\uvec k}_\perp,\Delta,n_-)
\nonumber\\
&\overset{|\bar{\uvec k}_\perp|\gg\Lambda_\text{QCD}}{=}
-i\,4\pi\alpha_S\int \frac{\ud^{4}l}{(2\pi)^4}\int \ud \bar k^-\,\frac{C_F\,I^{[\Gamma]}_q+T_R\,I^{[\Gamma]}_g}{\left[(\bar l-\bar k)^{2}+i\epsilon\right]\left[k'^{2}+i\epsilon\right]\left[k^{2}+i\epsilon\right]}+\mathcal{O}(\alpha_{s}^{2}),\label{eq:largekTGTMD1}
\end{align}
where
\begin{align}
I^{[\Gamma]}_q&=\tfrac{1}{4}\sum_jd_{\mu\nu}(\bar l-\bar k,n_-)\,\Tr\!\left[\gamma^{\nu}\uslash k'\Gamma\uslash k\gamma^{\mu}\Gamma_j\right]W_{\Lambda'\Lambda}^{[\Gamma^j]}(P,\bar l,\Delta,n_-),\\
I^{[\Gamma]}_g&=\tfrac{1}{2}\,\Tr\!\left[\uslash k \gamma_{\sigma}(\uslash \bar k-\uslash l) \gamma_{\nu} \uslash k'\Gamma\right]\frac{1}{l'^{+} \, l^{+}}\,W_{\Lambda'\Lambda}^{+\nu;+\sigma}(P,\bar l,\Delta,n_-),
\end{align}
with $l'=\bar l+\tfrac{\Delta}{2}$ ($k'=\bar k+\tfrac{\Delta}{2}$), $l=\bar l-\tfrac{\Delta}{2}$ ($k=\bar k-\tfrac{\Delta}{2}$), and the polarization sum $d^{\mu\nu}$ from Eq.~(\ref{eq:polsum}).
We then apply the collinear approximation for the momentum $\bar{l}\approx[\tfrac{x}{z}P^{+},0,\uvec 0_\perp]$ in the hard scattering part, which can be done if we assume that the average proton momentum $P$ has a large plus-component $P^{+}$. 
Accordingly, we take into account only the leading chiral-even traces with $\Gamma^j=\left\{ \gamma^{+},\gamma^{+}\gamma_{5}\right\}$ and $\Gamma_j=\left\{ \gamma^{-},-\gamma^{-}\gamma_{5}\right\}$ in the Fierz transformation.
We skip the details of the calculation, which are similar to the ones in Ref.~\cite{Bacchetta:2008xw}, and directly give the results for the large-$\bar k_\perp$ behavior of $F^q_{1,4}$ and $G^q_{1,1}$ at $\xi=0$,
\begin{align}
F^q_{1,4}&=\frac{\alpha_S}{2\pi^{2}}\int_{x}^{1}\frac{\ud z}{z}\,\frac{M^{2}\left[C_{F}\,\tilde{H}^{q}(\tfrac{x}{z},0,-\uvec\Delta^2_\perp)-T_{R}\,(1-z)^{2}\,\tilde{H}^{g}(\tfrac{x}{z},0,-\uvec\Delta^2_\perp)\right]}{\left[\uvec k'^2_\perp+z(1-z)\tfrac{\uvec\Delta^2_\perp}{4}\right]\left[\uvec k^2_\perp+z(1-z)\tfrac{\uvec\Delta^2_\perp}{4}\right]},\label{pQCDFq14}\\
G^q_{1,1}&= -\frac{\alpha_S}{2\pi^{2}}\left\{\int_{x}^{1}\frac{\ud z}{z}\,\frac{M^{2}\left[ C_{F}\, H^{q}(\tfrac{x}{z},0,-\uvec\Delta^2_\perp)+ T_{R}\,(1-z)^{2}\, H^{g}(\tfrac{x}{z},0,-\uvec\Delta^2_\perp)\right]}{\left[\uvec k'^2_\perp+z(1-z)\tfrac{\uvec\Delta^2_\perp}{4}\right]\left[\uvec k^2_\perp+z(1-z)\tfrac{\uvec\Delta^2_\perp}{4}\right]}\right.\nonumber \\
 &\phantom{=-\frac{\alpha_S}{2\pi^{2}}}\left.\quad -\int_{x}^{1}\frac{\ud z}{z}\,\frac{\tfrac{\uvec\Delta^2_\perp}{4}\, T_{R}\,z(1-z)\,(2\tilde{H}_{T}^{g}+E_{T}^{g})(\tfrac{x}{z},0,-\uvec\Delta^2_\perp)}{\left[\uvec k'^2_\perp+z(1-z)\tfrac{\uvec\Delta^2_\perp}{4}\right]\left[\uvec k^2_\perp+z(1-z)\tfrac{\uvec\Delta^2_\perp}{4}\right]}\right\},\label{pQCDGq11}
\end{align}
with $T_{R} = \tfrac{1}{2}$.
Here we follow the GPD conventions of Ref.~\cite{Diehl:2003ny}.
\begin{figure}[t!]
\begin{center}
\epsfig{file=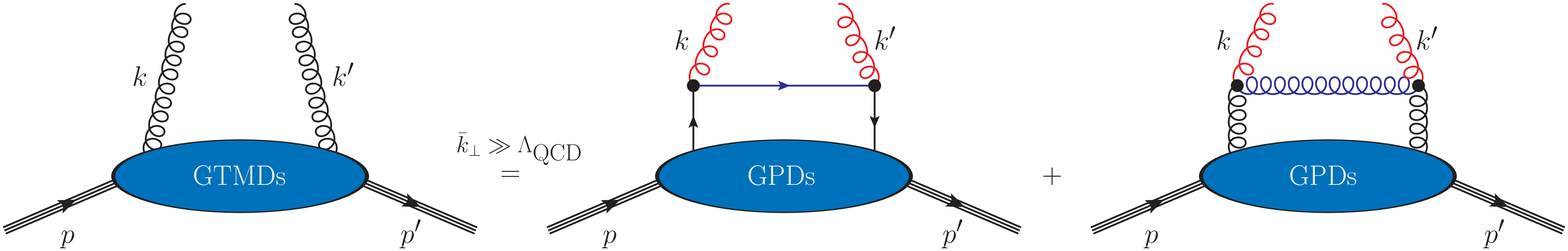,  width=1.0\textwidth}
\end{center}
\caption{\footnotesize{Diagrams determining the leading order of gluon GTMDs at large transverse momenta.}
\label{fig:GTMDgluon}}
\vspace{-0.2 truecm}
\end{figure}

Similarly, the large-$\bar k_\perp$ behavior of the gluon GTMDs is described perturbatively by the two diagrams on the right of Fig.~\ref{fig:GTMDgluon}. 
Expressing them by means of Feynman rules leads to the following formula,
\begin{align}
&\frac{1}{l'^{+} \, l^{+}}\,W_{\Lambda'\Lambda}^{+ \nu;+ \sigma}(P,x,\bar{\uvec k}_\perp,\Delta,n_-)\nonumber\\
&\overset{|\bar{\uvec k}_\perp|\gg\Lambda_\text{QCD}}{=}
-i\,4\pi\alpha_S\int \frac{\ud^{4}l}{(2\pi)^4}\int \ud \bar k^-\,\frac{C_F\,I^{\nu\sigma}_q+C_A\,I^{\nu\sigma}_g}{\left[(\bar l-\bar k)^{2}+i\epsilon\right]\left[k'^{2}+i\epsilon\right]\left[k^{2}+i\epsilon\right]}+\mathcal{O}(\alpha_{s}^{2}),\label{eq:largekTGTMD2}
\end{align}
where
\begin{align}
I^{\nu\sigma}_q&=\tfrac{1}{2}\sum_jd^{\nu\nu'}(k',n_-)\,d^{\sigma\sigma'}(k,n_-)\,\Tr\!\left[\gamma_{\nu'}(\uslash \bar k-\uslash \bar l)\gamma_{\sigma'}\Gamma_j\right]W_{\Lambda'\Lambda}^{[\Gamma^j]}(P,\bar l,\Delta,n_-),\\
I^{\nu\sigma}_g&=d^{\nu\nu'}(k',n_-)\, d^{\sigma'\sigma}(k,n_-)\, d^{\tau\tau'}(l-k,n_-)\,V_{1,\sigma'\tau\beta}\,V_{2,\nu'\tau'\alpha}\,\frac{1}{l'^{+} \, l^{+}}\,W_{\Lambda'\Lambda}^{+ \alpha;+ \beta}(P,\bar l,\Delta,n_-),
\end{align}
with 
\begin{align}
V_{1,\sigma'\tau\beta}&=g_{\sigma'\tau}(l-2k)_\beta+g_{\tau\beta}(k-2l)_{\sigma'}+g_{\beta\sigma'}(l+k)_{\tau},\\
V_{2,\nu'\tau'\alpha}&=g_{\tau'\nu'}(l'-2k')_\alpha+g_{\alpha\tau'}(k'-2l')_{\nu'}+g_{\nu'\alpha}(l'+k')_{\tau'}.
\end{align}
We then find for the large-$\bar k_\perp$ behavior of $F^g_{1,4}$ and $G^g_{1,1}$ at $\xi=0$ and for $x\geq 0$,
\begin{align}
F^g_{1,4}&=\frac{\alpha_S}{2\pi^{2}}\int_{x}^{1}\frac{\ud z}{z}\,\frac{M^{2}\left[C_{F}\,(1-z)\,\tilde{H}^{q}(\tfrac{x}{z},0,-\uvec\Delta^2_\perp) + C_A\,(z^2-4z+\tfrac{7}{2})\,\tilde{H}^{g}(\tfrac{x}{z},0,-\uvec\Delta^2_\perp)\right]}{\left[\uvec k'^2_\perp+z(1-z)\tfrac{\uvec\Delta^2_\perp}{4}\right]\left[\uvec k^2_\perp+z(1-z)\tfrac{\uvec\Delta^2_\perp}{4}\right]},\label{pQCDFg14}\\
G^g_{1,1}&= \frac{\alpha_S}{2\pi^{2}}\left\{\int_{x}^{1}\frac{\ud z}{z}\,\frac{M^{2}\left[C_{F}\,\tfrac{(1-z)(z-2)}{z}\,H^{q}(\tfrac{x}{z},0,-\uvec\Delta^2_\perp)+ C_A\,(2z^2-4z+\tfrac{7}{2}-\tfrac{2}{z})\, H^{g}(\tfrac{x}{z},0,-\uvec\Delta^2_\perp)\right]}{\left[\uvec k'^2_\perp+z(1-z)\tfrac{\uvec\Delta^2_\perp}{4}\right]\left[\uvec k^2_\perp+z(1-z)\tfrac{\uvec\Delta^2_\perp}{4}\right]}\right.\nonumber \\
 &\phantom{=\frac{\alpha_S}{2\pi^{2}}}\left.\quad +\int_{x}^{1}\frac{\ud z}{z}\,\frac{\tfrac{\uvec\Delta^2_\perp}{4}\, \tfrac{C_A}{2}\,(1-2z)^2\,(2\tilde{H}_{T}^{g}+E_{T}^{g})(\tfrac{x}{z},0,-\uvec\Delta^2_\perp)}{\left[\uvec k'^2_\perp+z(1-z)\tfrac{\uvec\Delta^2_\perp}{4}\right]\left[\uvec k^2_\perp+z(1-z)\tfrac{\uvec\Delta^2_\perp}{4}\right]}\right\},\label{pQCDGg11}
\end{align}
with $C_A = N_c = 3$.
The perturbative QCD results in~\eqref{pQCDFq14}, \eqref{pQCDGq11} for quarks and in~\eqref{pQCDFg14}, \eqref{pQCDGg11} for gluons clearly show that at large-$\bar k_\perp$ both $F_{1,4}$ and $G_{1,1}$ in general are nonzero.
This re-confirms in a model-independent way that these functions have to be included in the leading-twist parametrization of GTMDs.

%
%
\section{Summary}
\label{s:Summary}
In hadronic physics, GTMDs have already attracted considerable attention. 
Often they are denoted as {\it mother functions} because many GTMDs reduce to GPDs or TMDs in certain kinematical limits.
In this work we have focussed on two particular twist-2 GTMDs --- denoted by $F_{1,4}$ and $G_{1,1}$ in Ref.~\cite{Meissner:2009ww} --- which neither survive the GPD-limit nor the TMD-limit since in the decomposition of the GTMD correlator they are accompanied by a factor that is linear in both the transverse parton momentum and the transverse-momentum transfer to the nucleon.
In some sense this makes these two functions actually unique.
The particular role played by $F_{1,4}$ and $G_{1,1}$ is nicely reflected by their intimate connection to the orbital angular momentum of partons in a longitudinally polarized nucleon and in an unpolarized nucleon, respectively~\cite{Lorce:2011kd,Hatta:2011ku,Lorce:2011ni,Ji:2012sj,Lorce:2012ce,Lorce:2014mxa}.
Those relations open up new opportunities to study the spin/orbital structure of the nucleon.
For instance, new insights in this area could now be obtained through Lattice QCD, given the related pioneering and encouraging studies of TMDs and other parton correlation functions in Refs.~\cite{Hagler:2009mb,Musch:2010ka,Musch:2011er,Ji:2013dva}.

Our main intention in this paper has been to address recent criticism according to which even the mere existence of $F_{1,4}$ and $G_{1,1}$ is questionable~\cite{Liuti:2013cna,Courtoy:2013oaa}.
To this end we have shown in a model-independent way why the criticism of~\cite{Liuti:2013cna,Courtoy:2013oaa} does not hold.
We have also computed $F_{1,4}$ and $G_{1,1}$ in the scalar diquark model of the nucleon, in the quark-target model, and in perturbative QCD for large transverse parton momenta, and we generally have found nonzero results in lowest nontrivial order in perturbation theory.
Moreover, in the two spectator models we have verified explicitly the relation between the two GTMDs and the OAM of partons.
In summary, we hope that our work helps to resolve any potential doubt/confusion with regard to the status of the GTMDs $F_{1,4}$ and $G_{1,1}$ and their relation to the spin/orbital structure of the nucleon.

%
%
\begin{acknowledgments}
This work has been supported by the Grant-in-Aid for Scientific Research from the Japan Society of Promotion of Science under Contract No.~24.6959 (K.K.), the Belgian Fund F.R.S.-FNRS {\it via} the contract of Charg\'e de recherches (C.L.), the National Science Foundation under Contract No.~PHY-1205942 (A.M.), and the European Community Joint Research Activity ``Study of Strongly Interacting Matter'' (acronym HadronPhysics3, Grant Agreement No.~283286) under the Seventh Framework Programme of the European Community (B.P.).
The Feynman diagrams in this paper were drawn using JaxoDraw~\cite{Binosi:2008ig}.
\end{acknowledgments}


\end{document}